\documentclass[pdflatex,sn-mathphys]{sn-jnl}% Math and Physical Sciences Reference Style
\jyear{2022}%

\theoremstyle{thmstyleone}%
\usepackage{subcaption}
\usepackage{tablefootnote}
\theoremstyle{thmstyletwo}%

\theoremstyle{thmstylethree}%
\def\X{{\mathbf X}}

\def\ie{\textit{i}.\textit{e}., }
\def\eg{\textit{e}.\textit{g}., }

\newcommand{\SMV}{$\texttt{SARNet}_\texttt{MV}$}

\begin{document}

\title[THz Tomographic Imaging via Physics-Guided Restoration]{Making the Invisible Visible: Toward High-Quality Terahertz Tomographic Imaging via Physics-Guided Restoration}

%%=============================================================%%
%% Prefix	-> \pfx{Dr}
%% GivenName	-> \fnm{Joergen W.}
%% Particle	-> \spfx{van der} -> surname prefix
%% FamilyName	-> \sur{Ploeg}
%% Suffix	-> \sfx{IV}
%% NatureName	-> \tanm{Poet Laureate} -> Title after name
%% Degrees	-> \dgr{MSc, PhD}
%% \author*[1,2]{\pfx{Dr} \fnm{Joergen W.} \spfx{van der} \sur{Ploeg} \sfx{IV} \tanm{Poet Laureate} 
%%                 \dgr{MSc, PhD}}\email{iauthor@gmail.com}
%%=============================================================%%

\author[1]{\fnm{Weng-Tai} \sur{Su}}\email{wengtai2008@hotmail.com}

\author[1]{\fnm{Yi-Chun} \sur{Hung}}\email{nick831111@gmail.com}
% \equalcont{These authors contributed equally to this work.}

\author[1]{\fnm{Po-Jen} \sur{Yu}}\email{jerry321ab@gmail.com}
% \equalcont{These authors contributed equally to this work.}

\author[1]{\fnm{Shang-Hua} \sur{Yang}}\email{shanghua@ee.nthu.edu.tw}
% \equalcont{These authors contributed equally to this work.}

\author*[1]{\fnm{Chia-Wen} \sur{Lin}}\email{cwlin@ee.nthu.edu.tw}
% \equalcont{These authors contributed equally to this work.}
% Department
% of Electrical Engineering, National Tsing Hua University
% 101, Section 2, Kuang-Fu Road, Hsinchu 300044, Taiwan R.O.C.
\affil[1]{\orgdiv{Department
of Electrical Engineering}, \orgname{National Tsing Hua University}, \orgaddress{\street{Kuang-Fu Road}, \city{Hsinchu}, \postcode{30048},  \country{Taiwan}}}

%\affil[2]{\orgdiv{Department of Electrical and Computer Engineering}, \orgname{University of California, Los Angeles}, \orgaddress{\street{56-125B Engineering IV Building 420 Westwood Plaza}, \city{Los Angeles}, \postcode{90095-1594},  \country{USA}}}
% \affil[2]{\orgdiv{Department}, \orgname{Organization}, \orgaddress{\street{Street}, \city{City}, \postcode{10587}, \state{State}, \country{Country}}}

% \affil[3]{\orgdiv{Department}, \orgname{Organization}, \orgaddress{\street{Street}, \city{City}, \postcode{610101}, \state{State}, \country{Country}}}

%%==================================%%
%% sample for unstructured abstract %%
%%==================================%%

\abstract{Terahertz (THz) tomographic imaging has recently attracted significant attention thanks to its non-invasive, non-destructive, non-ionizing, material-classification, and ultra-fast nature for object exploration and inspection. However, its strong water absorption nature and low noise tolerance lead to undesired blurs and distortions of reconstructed THz images. The diffraction-limited THz signals highly constrain the performances of existing restoration methods.  To address the problem, we propose a novel multi-view Subspace-Attention-guided Restoration Network (SARNet) that  fuses  multi-view and multi-spectral features of  THz images for effective image restoration and 3D tomographic reconstruction. To this end, SARNet uses multi-scale branches to extract intra-view spatio-spectral amplitude and phase features  and fuse them via shared subspace projection and self-attention guidance. We then perform inter-view fusion to further improve the restoration of individual views by leveraging the redundancies between neighboring views. Here, we experimentally construct a THz time-domain spectroscopy (THz-TDS) system  covering a broad frequency range from 0.1 THz to 4 THz for building up a temporal/spectral/spatial/material  THz  database  of hidden 3D objects. Complementary to a quantitative evaluation, we demonstrate the effectiveness of our SARNet model on 3D THz tomographic reconstruction applications.}


%%\pacs[JEL Classification]{D8, H51}

%%\pacs[MSC Classification]{35A01, 65L10, 65L12, 65L20, 65L70}

\maketitle

\section{Introduction} 
\label{sec:intro}

Ever since the first camera's invention, imaging under different bands of electromagnetic (EM) waves, especially X-ray and visible lights, has revolutionized our daily lives~\cite{kamruzzaman2011application, rotermund1991methods, yujiri2003passive}. X-ray imaging plays a crucial role in medical diagnoses, such as cancer, odontopathy, and COVID-19 symptom~\cite{abbas2021classification, round2005preliminary, tuan2018dental}, based on X-ray's high penetration depth to great varieties of materials; visible-light imaging has not only changed the way of recording lives but contributes to the development of artificial intelligence (AI) applications, such as surveillance security and surface defect inspection \cite{xie2008review}. However, X-ray and visible-light imaging still face tough challenges. X-ray imaging is ionizing, which is harmful to biological objects and thus severely limits its application scope \cite{de2004risk}. On the other hand, although both non-ionizing and non-destructive, visible-light imaging cannot retrieve interior information of most objects which are opaque in visible light due to the highly absorptive and intense scattering behaviors between light and matter in the visible light band. To visualize the 3D information of objects in a remote but accurate manner, terahertz (THz) imaging has become among the most promising candidates among all EM wave-based imaging techniques
~\cite{abraham2010non, yu2012potential}.

%  \begin{table*}[!htb]
%      \centering
%      \caption{Comparison of features of existing imaging technologies. The ability to see through optically-opaque objects enables tomography. The X-ray would ionize objects, which means not bio-safe. Some methods can identify different materials by its spectroscopy, and it requires the penetration of the object. One imaging method is more favorable if it can be placed on the table (table-top), thereby excluding those methods which requires bulky instruments and placed in a special room such as X-ray and Magnetic Resonance Imaging (MRI).}
%      \vspace{0.1in}
%      % \label{tab:comparison}
%      \scalebox{0.8}{
%      \begin{tabular}{L{2.0cm} C{1.0cm} C{1.0cm} C{1.0cm} C{1.0cm}}
%      \hline
%      {\bf Method}         &  See through    &  Bio-safe &  Material   &  Table-top \\
%      &opaque objects & &Identification &system\\
%      \hline
%      \hline
%      Camera        &  $\times$  & $\checkmark$  & $\times$  & $\checkmark$ \\
%      X-ray         &  $\checkmark$  & $\times$  & $\checkmark$  & $\checkmark$ \\
%      LiDAR         &  $\times$  & $\checkmark$  & $\times$  & $\checkmark$ \\
%      Ultrasonic    &  $\checkmark$  & $\checkmark$  & $\times$  & $\checkmark$ \\
%      MRI           &  $\checkmark$  & $\checkmark$  & $\times$  & $\times$ \\
%      \hline
%      {\bf THz Imaging}   &  \textcolor{red}{$\checkmark$}  &  \textcolor{red}{$\checkmark$}  &  \textcolor{red}{$\checkmark$}  &  \textcolor{red}{$\checkmark$} \\
%      \hline
%      \end{tabular}}
%      \label{table:methods}
%  \end{table*}

\begin{table}[!htb]
    \centering
    \caption{Comparison of features of existing imaging technologies. The ability to see through objects opaque in visible light} enables tomography. The X-ray would ionize objects, which means not bio-safe. Some methods can identify different materials by their spectroscopy, and they require the penetration of the object. One imaging method is more favorable if it can be placed on the table (table-top), thereby excluding those methods which require bulky instruments and placed in a special room such as X-ray and Magnetic Resonance Imaging (MRI).
%    \vspace{0.1in}
    % \label{tab:comparison}
    \scalebox{0.75}{
    \begin{tabular}{c|c|c|c|c}%L{1.0cm} C{0.25cm} C{0.25cm} C{0.25cm} C{0.25cm}
    \hline
    {\bf Method}         &  See through    &  Bio-safe &  Material  &  Table-top \\
    &opaque objects & &Identification &system\\
    \hline
    \hline
      
        RGB Camera  &   $\times$ &   $\checkmark$  & Partially$^a$ & $\checkmark$  \\  
       
    \hline
    X-ray         &  $\checkmark$  & $\times$  & $\checkmark$  & $\checkmark$ \\
    \hline 
    LiDAR         &  $\times$  & $\checkmark$   & Partially$^b$ & $\checkmark$ \\
    \hline
    Ultrasonic    &  $\checkmark$  & $\checkmark$  & $\checkmark$  & $\checkmark$ \\
    \hline
    MRI           &  $\checkmark$  & $\checkmark$  & Partially$^c$  & $\times$ \\
    \hline
    {\bf THz Imaging}   &  \textcolor{red}{$\checkmark$}  &  \textcolor{red}{$\checkmark$}  &  \textcolor{red}{$\checkmark$}  &  \textcolor{red}{$\checkmark$}\\
    \hline
    \end{tabular}}\\
    \footnotesize{$^a$ Material of object surface (Fabric, plastic, wood, paper, leather, metal, and fur) \cite{kim2018rgbd}}\\
    \footnotesize{$^b$ Material of object surface (fabric, brick, pine, wood, and maple leaves) \cite{nunes2020polarization}}\\
    \footnotesize{$^c$ Material with hydrogen atoms (tumor, fat, and water) \cite{clarke1995mri}}
    \label{table:methods}
%    \vspace{-0.18in}
\end{table}

Table~\ref{table:methods} shows the comparison of different types of high-resolution imaging modalities with a non-contact setting. As camera and Light Detection and Ranging (LiDAR) are widely launched for 2D/3D image capturing, due to the intensive scattering and absorption happening nearby object surfaces, these two imaging methods cannot visualize 3-D full profiles of most objects. Research groups have successfully found other electromagnetic spectrum regimes to bring information invisible to visible to address this issue. X-ray imaging is one of the commonly used methods to precisely visualize the interior of objects \cite{chapman1997diffraction, fitzgerald2000phase, sakdinawat2010nanoscale, cloetens1996phase, peterson2001x}. Despite its invisible-to-visible capability, high-energy X-ray photons would cause both destructive and ionizing impacts on various material types preventing further investigations with other material characterization modalities. Magnetic resonance imaging (MRI) technology has proven to be a bio-safe way to visualize soft materials with excellent image contrast. Still, it is bulky and requires sufficient space for operation, which prevents its practical use in many application scenarios. To be pervasively used like visible light cameras, the desired tomographic imaging modality must be operated at a remote distance, non-destructive, bio-safe, compact, and most importantly, capable of digging out information conventional cameras cannot achieve.

% \begin{figure}[!hbt]
% \vspace{0.1in}
% \centering
% \includegraphics[width=0.5\textwidth]{figures/time_freq_v2.png}
% \caption{Raw data of measured THz images. This figure illustrates the time domain data measured in air and the body and leg of our 3D printed deer. The red points illustrate the frequency bands with strong water absorption. The right figures illustrate the reconstructed image using max value of time domain (upper right), and reconstructed image using different water absorption frequencies (lower right).}
% %\vspace{-0.25in}
% \label{fig:time_freq}	
% % \vspace{-0.15in}
% \end{figure}

%\vspace{-0.15in}
\begin{figure*}[!t]
\centering
\includegraphics[width=1.0\textwidth]{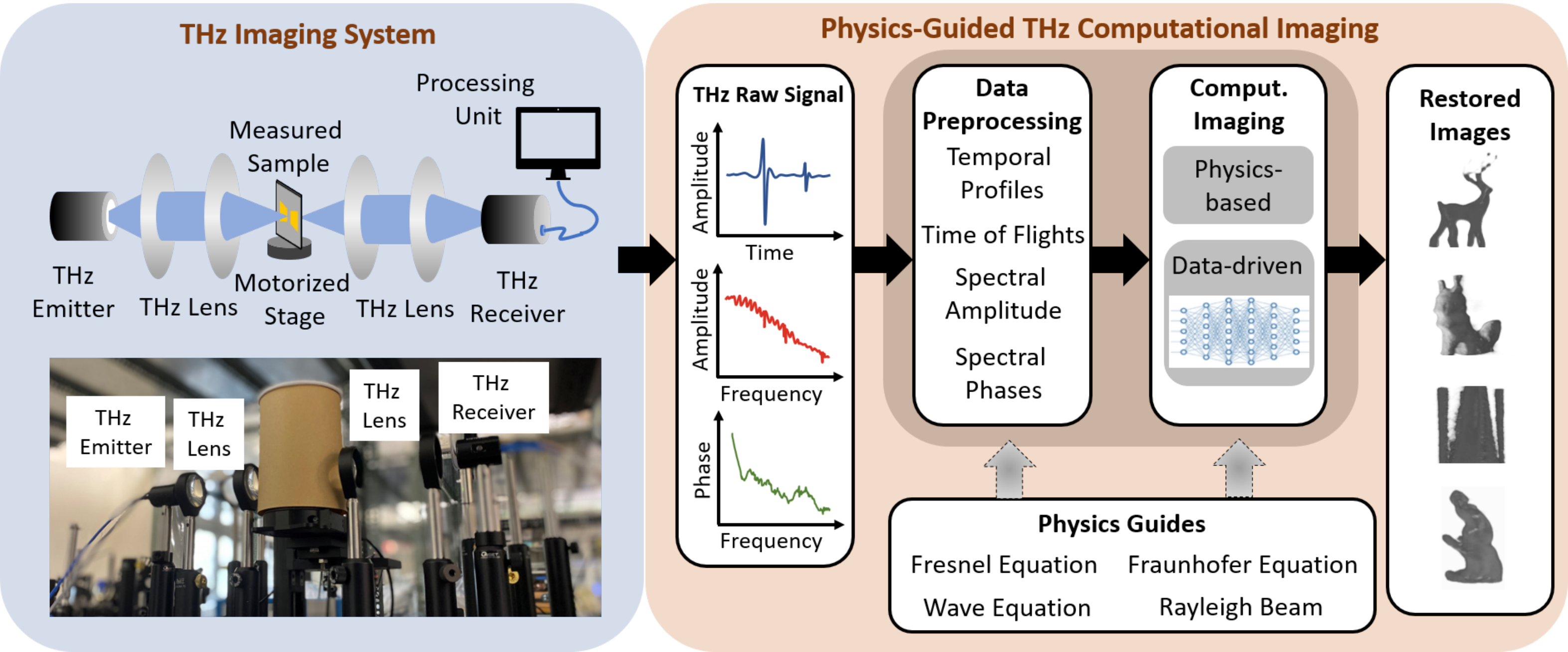}
%\vspace{-0.25in}
\caption{Flowchart of physics-guided THz computational imaging. The pixel-wise THz raw signals are measured from the THz imaging system along with image data. The multi-domain data are then processed and fused by a computational imaging model to reconstruct the images. The computational imaging model can be either physics-based or data-driven, with or without the physics guides derived from the physical properties of THz signals.} 
\label{fig:physical}	
% \vspace{-0.25in}
\end{figure*}

THz radiation, between microwave and infrared, has often been regarded as the last frontier of EM wave~\cite{saeedkia2013handbook}, which provides its unique functionalities among all EM bands. Along with the rapid development of THz technology, THz imaging has recently attracted significant attention due to its non-invasive, non-destructive, non-ionizing, material-classification, and ultra-fast nature for advanced material exploration and engineering. As THz waves can partially penetrate through varieties of materials being opaque in visible light, they carry hidden material tomographic information along the traveling path, making this approach a desired way to see through black boxes without damaging the exterior \cite{mittleman1999recent, jansen2010terahertz, mittleman2018twenty}. By utilizing light-matter interaction within the THz band, multifunctional tomographic information from a great variety of materials can also be retrieved even at a remote distance. In the past decades, THz time-domain spectroscopy (THz-TDS) has become one of the most representative THz imaging modalities to achieve non-invasive evaluation because of its unique capability of extracting geometric and multi-functional information of objects. Owing to its fruitful information in multi-dimensional domains --- space, time, frequency, and phase, THz-TDS imaging has been already allocated for numerous emerging fields, including drug detection~\cite{kawase2003non}, industrial inspection, cultural heritage inspection~\cite{fukunaga2016thz}, and cancer detection~\cite{bowman2018pulsed}.

However, the conventional methods (\eg \texttt{Time-max}~\cite{hung2019terahertz}) for THz imaging is to analyze the temporal profiles of THz signals measured by a THz-TDS system within a limited time window. The reconstructed tomographic image quality is severely constrained by the diffraction-limited geometry and absorption behavior of objects in the THz spectral regime, leading to undesired blurring and distortion of reconstructed tomography images. To address this problem, we utilize useful spectral bands to supplement the conventional method, recording the maximum amplitude of the time-domain THz signal of each pixel for recovery of the clear 2D images.

Recently, data-driven methods based on deep learning models, which do not resort to any explicit transform model but are learned from representative big data, have been revolutionizing the physics-based paradigm in image restoration. The data-driven methods can be regardless of physical properties while maintaining the advantages of the physics-based methods and achieving state-of-the-art performances.  We can also cast THz image analysis as an image-domain learning problem.  Nevertheless, a THz image retrieved from THz raw time-domain signals does not carry enough restoration information, thereby limiting the efficacy of the data-driven methods. Furthermore, we found that directly learning from the \textbf{full spectral information} to restore THz images leads to unsatisfactory performance. The main reason is that the full spectra of THz signals involve diverse characteristics of materials, noises, and scattered signals, which causes difficulties in model training. To address the above issues, as illustrated in Fig.~\ref{fig:physical}, we can leverage additional pixel-wise spectral information carried in the THz raw signals, such as the amplitude/phase spectra corresponding to specific physical characteristics of THz waves passing through materials. Due to a large number of spectral bands with measured THz image data, it is desirable to sample a subset of the most physics-prominent spectral bands to reduce the number of training parameters. Specifically, The THz beam is significantly attenuated at water absorption frequencies. As a result, such physics-guided water-absorption property of THz beams offers useful clues for inspecting and reconstructing an object from THz images captured in a see-through setting (\eg computed tomographic reconstruction) as will be elaborated in Sec.~\ref{sec:properties}.

Based on the concept revealed in Fig.~\ref{fig:physical}, we here propose a multi-scale \textbf{S}ubspace-\textbf{A}ttention guided \textbf{R}estoration \textbf{Net}work ($\texttt{SARNet}$) that fuses intra-view complementary spectral features of the THz amplitude and phase to supplement the \texttt{Time-max} image for restoring clear 2D images. To this end, \texttt{SARNet} learns common representations in a common latent subspace shared between the amplitude and phase, and then incorporates a Self-Attention mechanism to learn the wide-range dependency of the spectral features for guiding the restoration task. To leverage the inter-view redundancies existing between neighboring views of an object captured from different angles, on top of $\texttt{SARNet}$ we also propose a multi-view version image restoration model, namely $\texttt{SARNet}_\texttt{MV}$, that incorporates inter-view fusion to further boost restoration performance. Finally, from clear 2D views restored from the corrupted views of an object, we can reconstruct high-quality 3D tomography via inverse Radon transform. Our main contributions are summarized as follows:

\begin{itemize}%项目符号开始
\item We are the first research group to merge THz temporal-spatial-spectral data, data-driven models, and light-matter interaction properties to the best of our knowledge. The proposed \texttt{SARNet} achieves excellent performance in extracting and fusing features from the light-matter interaction data in THz spectral regime, which inherently contains fruitful 3D object information and its material behaviors. Based on the architecture of the proposed $\texttt{SARNet}_\texttt{MV}$ on intra/inter-view feature fusion, it delivers state-of-the-art performance on THz image restoration.

%We are the first to merge THz spatio-spectral data, data-driven models, and material properties. This work bridges the THz world and computer vision community, which opens up a new interdisciplinary research field to facilitate many practical applications with THz imaging, e.g., non-invasive evaluation, industrial inspection, gas tomography, and defect inspection.
%\item An efficient subspace and self-attention fusion Unet, namely SARNet, is proposed to deal with the image restoration task. By fusing and extracting the multi-spectral features in our proposed network architecture, SARNet significantly outperforms state-of-the-art algorithms in terms of restoration performance.

%We experimentally construct an ultra-fast, high SNR THz-TDS system covering a broad frequency range from 0.1 THz to 4 THz. This customized tool is designed to build up temporal/spectral/spatial/phase/material THz database of hidden 3D objects for development of THz imaging.

\item \noindent With our newly established THz-TDS tomography dataset --- the world's first in its kind, we provide comprehensive quantitative/qualitative analyses among $\texttt{SARNet}_\texttt{MV}$ and state-of-the-arts. $\texttt{SARNet}_\texttt{MV}$ significantly outperforms $\texttt{Time-max}$~\cite{hung2019terahertz}, $\texttt{U-Net}$~\cite{ronneberger2015u}, and $\texttt{NBNet}$~\cite{cheng2021nbnet} by 11.41 dB, 2.79 dB, and 2.23 dB, respectively, in average PSNR at reasonable computation and memory costs.

\item  This work shows that computer vision techniques can significantly contribute to the THz community and further open up a new interdisciplinary research field to boost practical applications, e.g., non-invasive evaluation, gas tomography,  industrial inspection, material exploration, and biomedical imaging.
\end{itemize}

% \vspace{-0.15in}
% \begin{figure}[t]
% \centering
% \includegraphics[width=0.5\textwidth]{figures/framework_v2.png}
% %\vspace{-0.25in}
% \caption{Illustration of a THz tomographic system framework composed of a THz material exploration system.} 
% \label{fig:flow}	
% \vspace{-0.25in}
% \end{figure}

\section{Related Work} 
\label{sec:related}

%In this section, we review works related to our research. First, we review some tomographic image
%reconstruction for X-ray and computed tomography (CT). Then, we review some deep learning-based restoration. 

%While there are some research works on image-based terahertz imaging \cite{popescu2010point, popescu2009phantom, wong2019computational}, till now there are no overview papers on terahertz imaging, especially on terahertz tomographic imaging for advanced materials exploration, from a signal processing perspective. In particular, machine learning-based image processing  can well exploit the rich spectral  information of terahertz images to facilitate the restoration and reconstruction of terahertz images. By focusing exclusively on the learning-based image processing  for terahertz materials exploration, this comprehensive overview paper will provide further depth to an important and practically useful application domain of terahertz materials exploration.  As terahertz imaging devices and getting mature and popular, terahertz tomographic imaging has been a powerful tool for advanced materials inspection and exploration. Image processing and computer vision techniques can significantly benefit the growth of this research field. It is thus at a good timing to provide a systematic overview of image processing-based terahertz tomographic imaging, and open issues that remain to be addressed in the future.

\subsection{Conventional THz Computational Imaging}
In the past decades, many imaging methods have been developed based on the light-matter interaction in the THz frequency range. Based on THz absorption imaging modalities, the material refractive index mapping can be profiled through Fresnel equation \cite{born2013principles}, extracted by the THz power loss while propagating through the tested object boundary. With THz spectroscopy imaging, both material information encoded in the wave propagation equation and object geometry can be revealed. 
To be more specific, the depth map of the measured object can be reconstructed based on the phase spectrum of the retrieved THz signals \cite{hack2014terahertz}; the attenuated power spectrum information can further recover the hyperspectral material fingerprint mapping. 
These characteristics provide functional 3D imaging capability for object inspection. 
Additionally, considering the propagated THz beam behavior of a signal as the model prior knowledge, such as Rayleigh beam, has proven to largely improve the imaging quality \cite{recur2012propagation}. 
With the THz time-reversal techniques, the THz amplitude and/or phase images of a measured object can be estimated by the spatiotemporal interaction between the input THz waves and the object.
However, the application scopes of those physics-driven methods are severely limited since they normally require a sufficient amount of prior knowledge of a measured object to simplify the guided complex physical models. To break this limitation, data-driven approaches, especially deep neural networks, start to arouse intensive attention due to their excellent learning capability. A data-driven model based on physical priors can effectively loosen the requirement of prior knowledge of materials and perform superior to conventional physics-based methods. Moreover, data-driven models can learn to adequately fuse the different information of THz signals, such as amplitude/phase spectra and the time-resolved THz signals, to achieve superior image restoration \cite{su2021seeing,su2023THzImaging}.

\subsection{Physics-guided Data-driven THz Imaging}
\label{sec:ip}
In contrast to those model-based methods, data-driven methods are mainly based on deep learning models \cite{zhang2017beyond, mao2016image}, which do not resort to any explicit transform model but are learned from representative big data. We can cast THz image analysis as an image-domain learning problem. Deep learning has revolutionized the aforementioned physics-based paradigm in image restoration, for which the data-driven methods can be regardless of physical properties while maintaining the advantages of the physics-based methods and achieving state-of-the-art performances. Nevertheless, a THz image retrieved from THz raw time-domain signals does not carry enough restoration information, thereby limiting the efficacy of the data-driven methods. To address the issue, as illustrated in Fig.~\ref{fig:physical}, we can leverage additional pixel-wise spectral information carried in the THz raw signals, such as the amplitude/phase spectra corresponding to specific physical characteristics of THz waves passing through materials. By contrast, the physics-based methods are difficult to leverage such pixel-wise amplitude/phase spectral information. To this end, the data-driven model proposed in \cite{su2021seeing} incorporates additional information from amplitude/phase at water absorption frequencies, derived from the physical properties of THz signals, to complement the insufficient information in time-domain THz images so as to significantly boost restoration performance. In addition, if the THz imaging system uses the THz focal beam, the THz beam diameter along with the wave propagation direction can be varied. Additionally, the THz beam diameter can also be changed in different spectral bands due to the diffraction limit. Both changed THz beam diameters lead to the non-identical point spread function (PSF) in each measurement point. To solve this problem, the Filter Adaptive Convolutional Layer (FAC) \cite{zhou2019spatio} can learn different filter kernels corresponding to the PSF for each pixel from spatial-spectral information and use those kernels to deliver superior imaging performance.

\subsection{Deep Learning-based Image Restoration}
In recent years, deep learning methods were first popularized in high-level visual tasks, and then gradually penetrated into many tasks such as image restoration and segmentation. 
%For example, \cite{chen2017low} proposed an encoder–decoder network to compress the input data into the latent representation, and then the decompression reconstruct the output data into a complete dataset, which had been mainly used to restore full dose CT images from low dose images. In PET/MR applications,  \cite{liu2018deep} employ attenuation correction and MR images to generate synthetic CT images. \cite{yang2017enhancing} proposed to use neural network (NN) to estimate the fusion parameters according to the maximum a posteriori (MAP) with different regularization weights for PET restoration, and \cite{cui2017deep} used the maximum likelihood expectation maximization (MLEM) for dynamic PET imaging, a reconstruction framework based on stacked sparse autoencoders. 
%As one of the most basic image processing problems
Convolutional neural networks (CNNs) have  proven to achieve state-of-the-art performances in fundamental image restoration problems~\cite{zhang2017beyond, mao2016image, zhang2020residual, zhang2018ffdnet, ronneberger2015u}. Several network models for image restoration were proposed, such as U-Net \cite{ronneberger2015u}, hierarchical residual network \cite{mao2016image} and residual dense network \cite{zhang2020residual}. Notably, DnCNN \cite{zhang2017beyond} uses convolutions, BN, and ReLU to build 17-layer network for image restoration which was not only utilized for blind image denoising, but was also employed for image super-resolution and JPEG image deblocking. FFDNet \cite{zhang2018ffdnet} employs noise level maps as inputs and utilizes a single model to develop variants for solving problems with multiple noise levels. In \cite{mao2016image} a very deep residual encoding-decoding (RED) architecture was proposed  to solve the image restoration problem using skip connections. \cite{zhang2020residual} proposed a residual dense network (RDN), which maximizes the reusability of features by using residual learning and dense connections. NBNet \cite{cheng2021nbnet} employs subspace projection to transform learnable feature maps into the projection basis, and leverages non-local image information to restore local image details.
Similarly, the \texttt{Time-max} image obtained from a THz imaging system can be cast as an image-domain learning problem which was rarely studied due to the difficulties in THz image data collection. Research works on image-based THz imaging include \cite{popescu2010point, popescu2009phantom, wong2019computational}, and THz tomographic imaging works include \cite{hung2019terahertz, hung2019kernel}.

Transformer~\cite{vaswani2017attention}, a kind of self-attention mechanism for machine learning, was first proposed to largely boost the research in natural language processing. Recently, it has gained wide popularity in the computer vision community, such as image classification~\cite{dosovitskiy2020image, wu2020visual}, object detection~\cite{carion2020end, liu2018deep}, segmentation~\cite{wu2020visual}, which learns to focus on essential image regions by exploring the long-range dependencies among different regions. Transformer has also been introduced for image restoration~\cite{chen2021pre, cao2021video, wang2022uformer} due to its impressive performance. In \cite{chen2021pre}, a standard Transformer-based backbone model IPT was proposed to address various restoration problems, which relies on a large number of parameters (over 115.5M parameters), large-scale training datasets, and multi-task learning for achieving high restoration performances. Additionally, VSR-Transformer~\cite{cao2021video} first utilizes a CNN to extract visual features and then adopts a self-attention model to fuse features for video super-resolution Although transformer-based attention mechanisms have proven effective in boosting the performance of image restoration tasks, the performance gains of transformers come at the cost of significantly larger amounts of training data and computation.

%In this paper, we use deep learning-based image restoration to solve this problem, especially how to fuse and extract the rich spectral information of THz images to facilitate the restoration and reconstruction of THz images, especially how to fuse the amplitude and phase with different physical characteristics. Inspired by U-Net \cite{ronneberger2015u} for multi-scale feature extraction, our work extends the similar idea of extracting multi-scale features from both amplitude and phase. In order to fuse multi-scale and different spectral features more effectively, we proposed the subspace and self-attention fusion module, which can effectively Fusion of spatial spectral features to obtain better restoration results.

\subsection{Tomographic Reconstruction}
Computed tomographic (CT) imaging methods started from X-ray imaging, and many methods of THz imaging are similar to those of X-ray imaging. One of the first works to treat X-ray CT as an image-domain learning problem was \cite{kang2017deep}, which adopts CNN to refine tomographic images. In \cite{jin2017deep}, U-Net was used to refine image restoration with significantly improved performances.  \cite{zhu2018image} further projects sinograms measured directly from X-ray into higher-dimensional space and uses domain transfer to reconstruct images. The aforementioned works were specially designed for X-ray imaging. 

Hyperspectral imaging \cite{schultz2001hyperspectral, ozdemir2020deep, geladi2004hyperspectral} constitutes  image modalities other than THz imaging. Different from THz imaging, Hyperspectral imaging collects continuous spectral band information of the target sample. Typically, the frequency bands fall in the visible and infrared spectrum; hence, most hyperspectral imaging modalities can only observe the surface characteristics of targeted objects. Furthermore, although existing deep-based hyperspectral imaging works can learn spatio-spectral information from a considerable amount of spectral cube data, they mainly rely on the full spectral information to restore hyperspectral images. This would usually lead to unsatisfactory performance for THz imaging since the full spectral bands of THz signals involve diverse characteristics of materials, noises, and scattered signals, which causes difficulties in model training.

\section{Physics-Guided THz Imaging} 
\label{sec:THz_signal}

Based on the dependency between the amplitude of a temporal signal and THz electric field, in conventional THz imaging, the maximum peak of the signal ($\texttt{Time-max}$~\cite{hung2019terahertz}) is extracted as the feature for a voxel. The reconstructed image based on $\texttt{Time-max}$ features can deliver a great signal-to-noise ratio and a clear object contour. However, the conventional THz imaging based on $\texttt{Time-max}$ features suffers from several drawbacks, such as the undesired contour in the boundary region, the hollow in the body region, and the blurs in high spatial-frequency regions. To break this limitation, we utilize the spectral information of THz temporal signals to supplement the conventional method based on $\texttt{Time-max}$ features since the voxel of the material behaviors is encoded in both the phase and amplitude of different frequency components, according to the Fresnel equation~\cite{dorney2001material}.

\begin{figure*}[t]
\vspace{0.1in}
\centering
\includegraphics[width=1\textwidth]{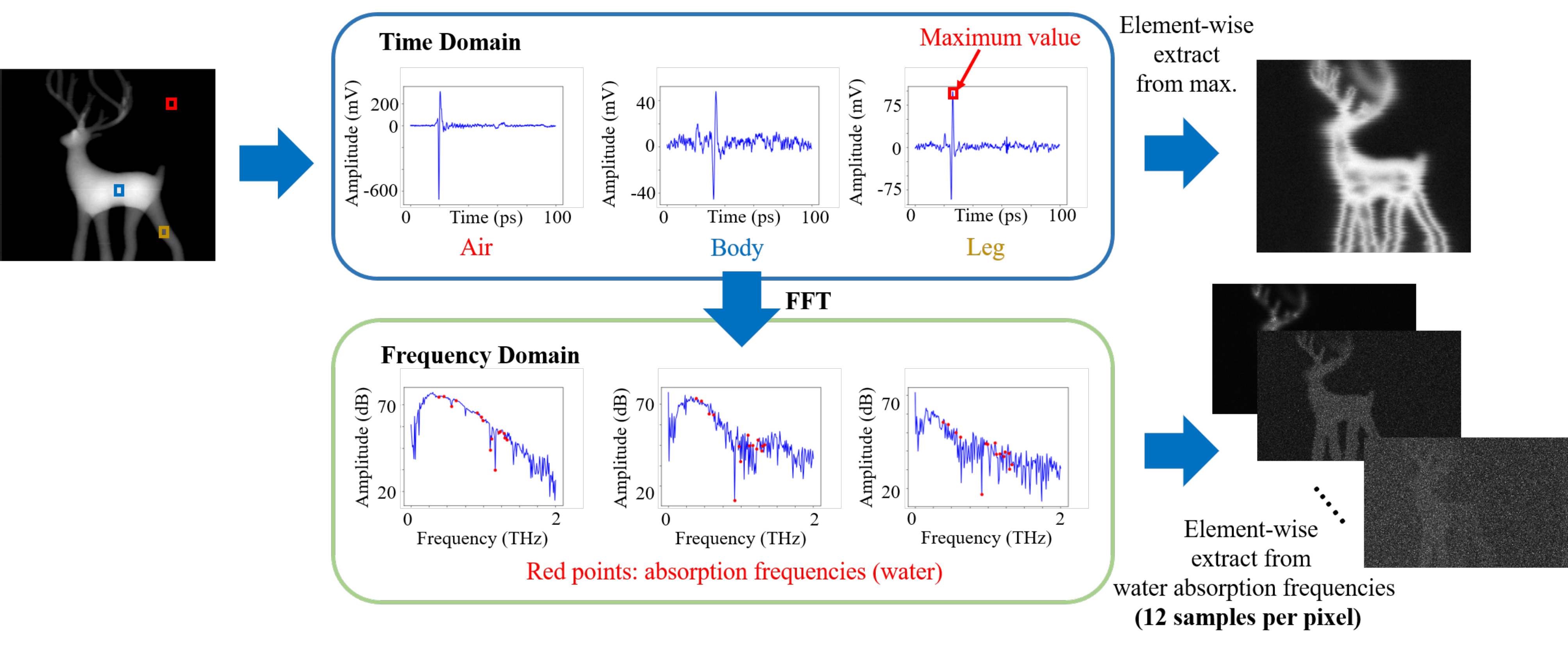}
\caption{Raw data of measured THz images. This figure illustrates the time domain data measured in air and the body and leg of our 3-D printed deer. The red points illustrate the frequency bands with strong water absorption. The right figures illustrate the reconstructed image using the max value of the time domain (upper right), and the reconstructed image using different water absorption frequencies (lower right).}
%\vspace{-0.25in}
\label{fig:time_freq}	
% \vspace{-0.15in}
\end{figure*}

More specifically, considering an incident THz wave penetrates through a single-material object with thickness $d$, the detected THz signal $S_\mathrm{d}(f)$ at frequency $f$ is determined by the material complex refractive index $\Tilde{n}_o(f)=n_o(f)-j \kappa_o(f)$ and the thickness $d$ in \eqref{eq:material abs}.

%\textcolor{blue}{More specifically, the detected power $S_{d}(f)$ of THz waves passing through an object is determined by the absorption coefficient $\alpha(f)$ and the thickness $L$ of the object in \eqref{eq:material abs}.} 

\begin{equation} \label{eq:material abs}
    \begin{aligned}
    S_d(f)&=S_\text{ref}(f)\cdot t(\Tilde{n}_o, f) \cdot \exp\left[\frac{\kappa_o(f)2\pi f d}{c}\right]\cdot \exp\left[\frac{-jn_o(f)2\pi f d}{c}\right] \\
    &= S_\text{ref}(f)\cdot t(\Tilde{n}_o, f) \cdot (I_{a}(f))^d \cdot (I_{p}(f))^d,
    \end{aligned}
\end{equation}
where $S_\mathrm{ref}(f)$ and $t(\Tilde{n}_o, f)$ are respectively the THz input signal and the Fresnel loss of THz waves (\textit{e.g.}, amplitude attenuation and phase change) due to the air-object interface at frequency $f$. 
Here, the Fresnel loss resulting from the presence of a single material can be further simplified as a constant. 
Meanwhile, $I_{a}(f)=\exp\left[\frac{\kappa_o(f)2\pi f}{c}\right]$ and $I_{p}(f)=\exp\left[\frac{-jn_o(f)2\pi f}{c}\right]$ can be acquired in a data-driven manner using information regarding the object thickness (\textit{i.e.}, ground-truth) and the detected THz signal. 
Specifically, although the complex refractive index is not provided explicitly, the network can still learn to map noisy input amplitude/phase images to their corresponding ground-truth images.
% \textcolor{blue}{where the interface loss $t(\Tilde{n}_H, f)$ caused by HIPS objects can be viewed as a constant, $I_{a}(f)=\exp\left[\frac{\kappa_H(f)2\pi f}{c}\right]$ and $I_{p}(f)=\exp\left[\frac{-jn_H(f)2\pi f}{c}\right]$ can be learned from the information of object thickness (\textit{i.e.}, the ground-truth) and the detected THz signal in a data-driven manner.  
% %
% Specifically, even though the complex refractive index is not explicitly given, the network still can learn the mapping between noisy input amplitude/phase images and their corresponding ground-truth images.}

To provide a more detailed explanation of THz imaging, Fig.~\ref{fig:time_freq} shows the flowchart of estimating amplitude and phase information of $S_d(f)$ from the raw data directly measured by the THz-TDS system. This figure illustrates time-domain THz signals measured in air, the body, and the leg of a 3-D printed deer, respectively. While the THz beam passes through the object, the attenuated THz time-domain signal encodes the thickness and material information of the THz-illuminated region. By processing the peak amplitudes of THz signals (\textit{i.e.,} \texttt{Time-max}), the 3-D profile of the printed deer can be further reconstructed. Although this conventional way is well-fitted for visualizing 3-D objects, the inherent diffraction behavior and strong water absorption nature of THz wave induce various kinds of noise sources as well as the loss of material information, as characterized by parameters such as  $t(\Tilde{n}_o, f)$, $I_{a}(f)$, and $I_{p}(f)$ in 
 \eqref{eq:material abs}. This leads to the undesirable blurring, distorted, speckled phenomenon of functional THz images. Existing works have tackled this issue to restore clear images via estimating point spread functions \cite{popescu2010point, popescu2009phantom}, image enhancement \cite{wong2019computational}, machine learning \cite{ljubenovic2020cnn, wong2019training}, and more. Their performance is, however, still severely constrained by diffraction-limited THz beam. To break the limitations, the motivation of our work is to reconstruct deep-subwavelength tomographic images by using a deep-learning-based image restoration method and spatio-spectral information of the hidden objects. 

\begin{figure*}
\centering
\includegraphics[width=1.0\textwidth]{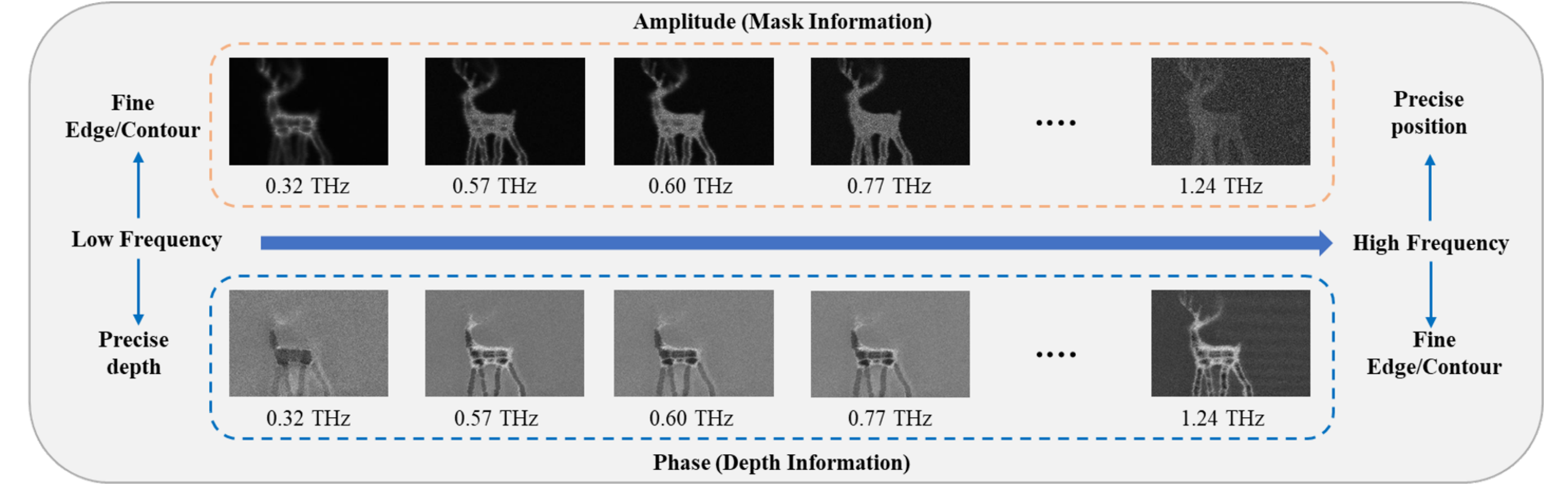}
\caption{Illustration of THz multi-spectral amplitude and phase images measured from \textbf{Deer}.} 
\label{fig:band}	
\end{figure*}

\subsection{Water Absorption Profile-guided THz Imaging}
\label{sec:properties_water}

% As shown in Fig. \ref{fig:time_freq}, each 2-D THz image is composed of an array of time-domain pulse signals with a temporal resolution of 10 fs. Through Fourier transform operation, we can extract pixel-wise multi-spectral features from the distributed time-domain THz signals.
% Due to the large number of spectral bands with measured THz image data, it is required to sample a subset of the spectral bands to reduce the training parameters. The THz beam is significantly attenuated at water absorption frequencies. Thus, reconstructed THz images based on water absorption lines offer worse details. Besides, the high-speed THz-TDS system offers more than 20 dB SNR in a frequency range of 0.3 THz--1.3 THz. Considering the water absorption in THz regime \cite{van1989terahertz, slocum2013atmospheric} and the superior SNR in the range of 0.3 THz--1.3 THz, we select 12 frequencies at 0.380, 0.448, 0.557, 0.621, 0.916, 0.970, 0.988, 1.097, 1.113, 1.163, 1.208, and 1.229 THz.
As shown in Fig.~\ref{fig:time_freq}, each 2-D THz image is composed of an array of time-domain signals, from which the  Fourier transform operation can be utilized to extract voxel-wise multi-spectral features.
Due to a large number of spectral bands with measured THz image data, it is required to sample a small subset of prominent spectral bands to reduce the training burden. Because the THz wave is significantly attenuated at water absorption frequencies, selecting THz bands based on water absorption lines can better delineate an object and characterize its thickness profile. %Besides, the high-speed THz-TDS system offers more SNR in a frequency range sub-THz to few THz. 
The spectral information, including both amplitude and phase at the selected frequencies, is extracted and then employed to restore clear 2D images. The different features in THz images at THz water-absorption frequencies (the 12 selected frequencies in this work: 0.380, 0.448, 0.557, 0.621, 0.916, 0.970, 0.988, 1.097, 1.113, 1.163, 1.208, and 1.229 THz) as shown in Fig.~\ref{fig:band}. It shows multiple 2D THz images of the same object at the selected frequencies, showing very different contrasts and spatial resolutions as these hyperspectral THz image sets have different physical characteristics through the interaction of THz waves with objects.

% Specifically, Fig.~\ref{fig:band} shows multiple 2D THz images of the same object at the selected frequencies, showing very different contrasts and spatial resolutions as these hyperspectral THz image sets have different physical characteristics through the interaction of THz waves with objects. 
%The phase contains more accurate depth information and resolution, but also contains the complex information of light-matter interaction that could cause the learning difficulties in the image restoration task. 
The lower-frequency phase images offer relatively accurate depth information due to their higher SNR level, whereas the higher-frequency phase images offer finer contours and edges because of the shrinking diffraction-limited wavelength sizes (from left to right in Fig.~\ref{fig:band}). The phase also contains, however, a great variety of information on light-matter interaction that could cause learning difficulty for the image restoration task. 
%
%Because the phase contains too many mixed physical quantities and is not easy to separate, this leads to poor reconstruction results. 
To address this issue, we utilize the amplitude spectrum as complementary information. Although the attenuated amplitude spectrum cannot reflect comparable depth accuracy levels as the phase spectrum, the amplitude spectrum still presents superior SNR and more faithful contours such as the location information of a measured object. Besides, as the complementary information to phase, the lower-frequency amplitude offers higher contrast, whereas the higher-frequency amplitude offers a better object mask.

In summary, the amplitude complements the shortcomings of the phase. The advantages of fusing the two signals from low-frequency to high-frequency are as follows: Since the low-frequency THz signal provides precise depth (the thickness of an object) and fine edge/contour information in the phase and amplitude, respectively, they together better delineate and restore the object. In contrast, the high-frequency feature maps of amplitude and phase respectively provide better edges/contours and precise position information, thereby constituting a better object mask from the complementary features. With these multi-spectral properties of THz images, we can extract rich information from a wide spectral range in the frequency domain to simultaneously restore the 2D THz images without any additional computational cost or equipment, which is beneficial for the further development of THz imaging.

\section{THz Tomographic Imaging} 
\label{sec:method}

\begin{figure*}[t]
\centering
\includegraphics[width=1\textwidth]{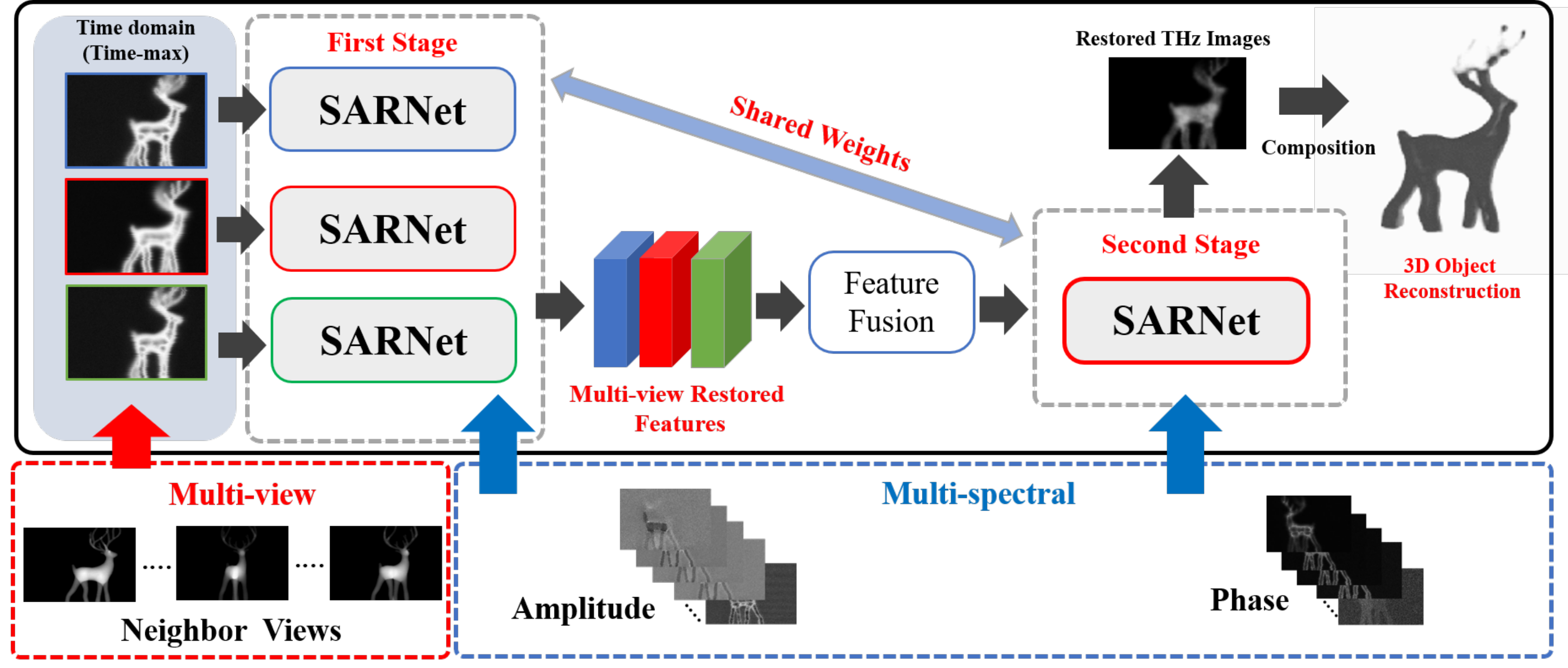}
%\vspace{-0.25in}
\caption{Illustration of THz 3D tomographic imaging based on \texttt{SARNet}.} 
\label{fig:overall}	
% \vspace{-0.2in}
\end{figure*}

\subsection{Overview}
\label{sec:overview}

As different EM bands interact with objects differently, THz waves can partially penetrate through various optically opaque materials and carry hidden material tomographic information along the traveling path. This unique feature provides a new approach to visualizing the essence of 3D objects, which other imaging modalities cannot achieve. Although existing deep neural networks can learn spatio-spectral information from a considerable amount of spectral cube data, as mentioned above, directly learning from the \textbf{full spectral information} is not appropriate for learning THz image restoration models since the full spectral bands of THz signals involve diverse characteristics of materials, noises, and scattered signal, which causes difficulties in model training. 

To address this problem, our work is based on extracting \textbf{complementary information} from both the amplitude and phase of a THz signal. In addition, for tomographic reconstruction, we capture  multi-view images of an object with overlapping contents from different view angles. As a result, the redundancies existing between neighboring views offer useful aid in enhancing the restoration qualities of individual corrupted views. In summary, we devise a novel multi-view Subspace-Attention-guided Restoration Network ($\texttt{SARNet}_\texttt{MV}$), as shown in Fig.~\ref{fig:overall}, to capture intra-view complementary spectral characteristics of materials and inter-view  redundancies from neighboring views to restore corrupted 2D THz images effectively. The key idea of $\texttt{SARNet}_\texttt{MV}$ is to fuse spatio-spectral features with different characteristics on a common ground via deriving a shared latent subspace and discovering the wide-range dependencies between the amplitude and phase images to guide the feature fusion. To this end, $\texttt{SARNet}_\texttt{MV}$ is a two-stage multi-view version based on single-view  $\texttt{SARNet}$.  In the first single-view stage of $\texttt{SARNet}_\texttt{MV}$, as shown in Fig.~\ref{fig:overall}, all corrupted views are first restored by  $\texttt{SARNet}$ individually. Then, in the second multi-view stage,  we first concatenate and fuse the feature tensors of three restored neighboring views via a feature fusion module, and then feed the fused multi-view feature into  the same \texttt{SARNet} to obtain the final restored view. The design of $\texttt{SARNet}_\texttt{MV}$ is detailed in Sec.~\ref{sec:network}. 
\subsection{Network Architecture}
\label{sec:network}
\begin{figure*}[!]
\centering
\includegraphics[width=1\textwidth]{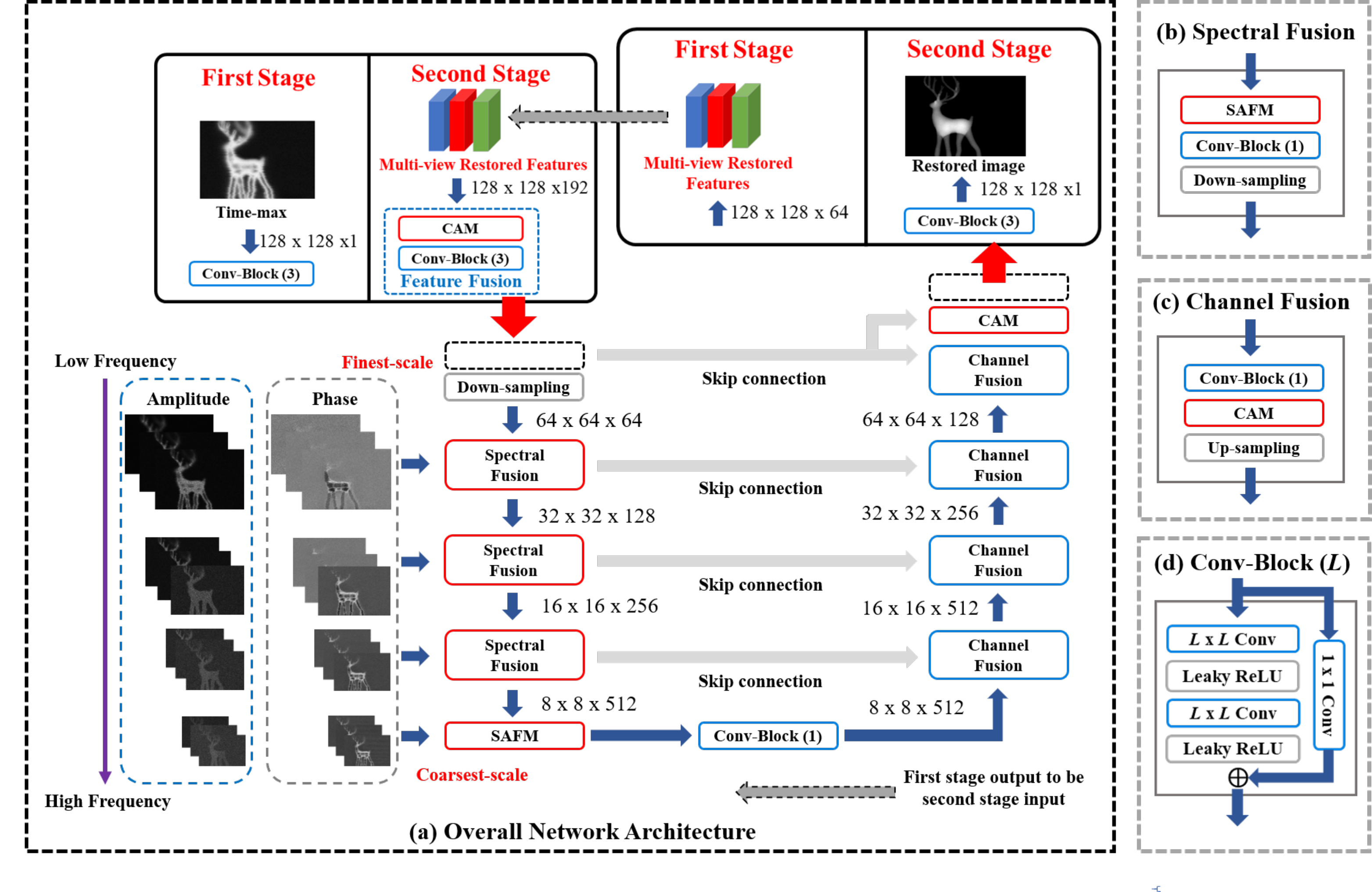}
% \vspace{-0.15in}
\caption{(a) Overall network architecture of $\texttt{SARNet}_\texttt{MV}$ consisting of five scale-branches, where the finest-scale scale takes the feature tensor of one view's \texttt{Time-max} image (the first stage) or the fused image of three restored views (the second stage) as input. Additionally, each of the second to fifth takes 6 images of spectral frequencies (\ie 3 amplitude bands and 3 phase bands) as inputs. The three gray blocks show the detailed structures of (b) Spectral Fusion, (c) Channel Fusion, and (d) Conv-Block. The two black blocks indicate the input and output in the first and second stages, respectively.} 
\label{fig:unet}	
% \vspace{-0.2in}
\end{figure*}

% \textcolor{red}{Furthermore, the low frequency mainly provides high contrast information, that is, the thickness information of the object, which approximate the depth map of the object. In the contrast, the high frequency mainly provides more detailed information, that is, the edge and position information of the object, which mean to provide the mask of object location.}
% Inspired by the CNN architecture from U-Net \cite{ronneberger2015u}, 
On top of \texttt{U-Net}~\cite{ronneberger2015u}, the architecture of \texttt{SARNet} is depicted in Fig.~\ref{fig:unet}.  Specifically, \texttt{SARNet} is composed of an encoder (spectral-fusion module) with 5 branches of different scales (from the finest to the coarsest) and a decoder (channel-fusion module) with 5 corresponding scale branches. Each scale branch of the encoder involves a Subspace-Attention-guided Fusion module (SAFM), a convolution block (Conv-block), and a down-sampler, except for the finest-scale branch that does not employ SAFM. To restore a specific view, the encoder of \texttt{SARNet} takes the feature tensor of this view's \texttt{Time-max} image (the first stage) or a fused image of three restored neighboring views centered at the current view (the second stage) as the input of the finest-scale branch. To extract and fuse multi-spectral features of both amplitude and phase in a multi-scale manner, the encoder also receives to its second to fifth scale branches 24 images of additional predominant spectral frequencies extracted from the THz signal of the current view, where each branch takes 6 images of different spectral bands (\ie 3 amplitude bands and 3 corresponding phase bands) to extract learnable features from these spectral bands. To reduce the number of model parameters, these 24 amplitude and phase images (from low to high frequencies) are downsampled to 4 different spatial scales and fed into the second to fifth scale branches in a fine-to-coarse manner as illustrated in Fig.~\ref{fig:unet}. We then fuse the multi-spectral amplitude and phase feature maps in each scale via the proposed SAFM that learns a common latent subspace shared between the amplitude and phase features to facilitate associating the self-attention-guided wide-range amplitude-phase dependencies. Projected into the shared latent subspace, the spectral features of amplitude and phase components, along with the down-sampled features of the upper layer, can then be properly fused together on common ground in a fine-to-coarse fashion to derive the final latent code. 

The Conv-block($L$) contains two stacks of $L \times L$ convolution, batch normalization, and ReLU operations. Because the properties of the spectral bands of amplitude and phase can be significantly different, we partly use $L=1$ to learn the best linear combination of multi-spectral features to avoid noise confusion and reduce the number of model parameters. The up-sampler and down-sampler perform $2\times$ and $\frac{1}{2} \times$ scaling, respectively. The skip connections~(SC) directly pass the feature maps of different spatial scales from individual encoder branches
to the Channel Attention Modules (CAMs) of their corresponding branches of the decoder. The details of SAFM and CAM are elaborated on later.

In the decoder path, each scale branch for channel fusion involves an up-sampler, a CAM, and a Conv-block. The Conv-block has the same functional blocks as that in the encoder. Each decoding branch receives a ``shallower-layer''  feature map from the corresponding encoding branch via the skip-connection shortcut and concatenates the feature map with the upsampled version of the decoded ``deeper-layer''  feature map from its coarser-scale branch.  Besides, the concatenated feature map is then processed by CAM to capture the cross-channel interaction to complement the local region for restoration. 

% For amplitude, the lower spectral-bands which mainly provide high-contrast features such as detail of edge and contour. In contrast, the higher spectral-bands which can obtain the semantic-level features of an object such as mask and position information. For phase, the lower spectral-bands which extract the precise depth to recovery the thickness of object pixels, carrying the depth information of the object. Then, the higher spectral-bands also provide the high contrast features of an object which are similar to the lower spectral-bands of amplitude.

Note, a finer-scale branch of \texttt{SARNet} extracts shallower-layer features that tend to capture low-level features, such as colors and edges. To complement the  \texttt{Time-max} image for restoration, we feed additional amplitude and phase images of low to high spectral bands into the fine- to coarse-scale branches of \texttt{SARNet}.  Since the spectral bands of THz amplitude and phase offer complementary information, as mentioned in Sec.~\ref{sec:overview}, besides the \texttt{Time-max} image \texttt{SARNet} also extracts multi-scale features from the amplitude and phase images of 12 selected THz spectral bands, which are then fused by the proposed SAFM.
% We feed the Time-max signal as the input of the finest-scale branch, then feed the 12 spectral bands to the remaining four branches (three bands to each scale): lower-frequency bands to finer-scale branches as shown in Fig.~ \ref{fig:unet}.

% \vspace{-0.02in}

% it is the best strategy to use the shallow and deep at the same time, while allowing the decoder to learn the relevant features that are lost in the encoder by using pooling down-sampling.

\begin{figure*}[t]
\centering
\includegraphics[width=1.05\textwidth]{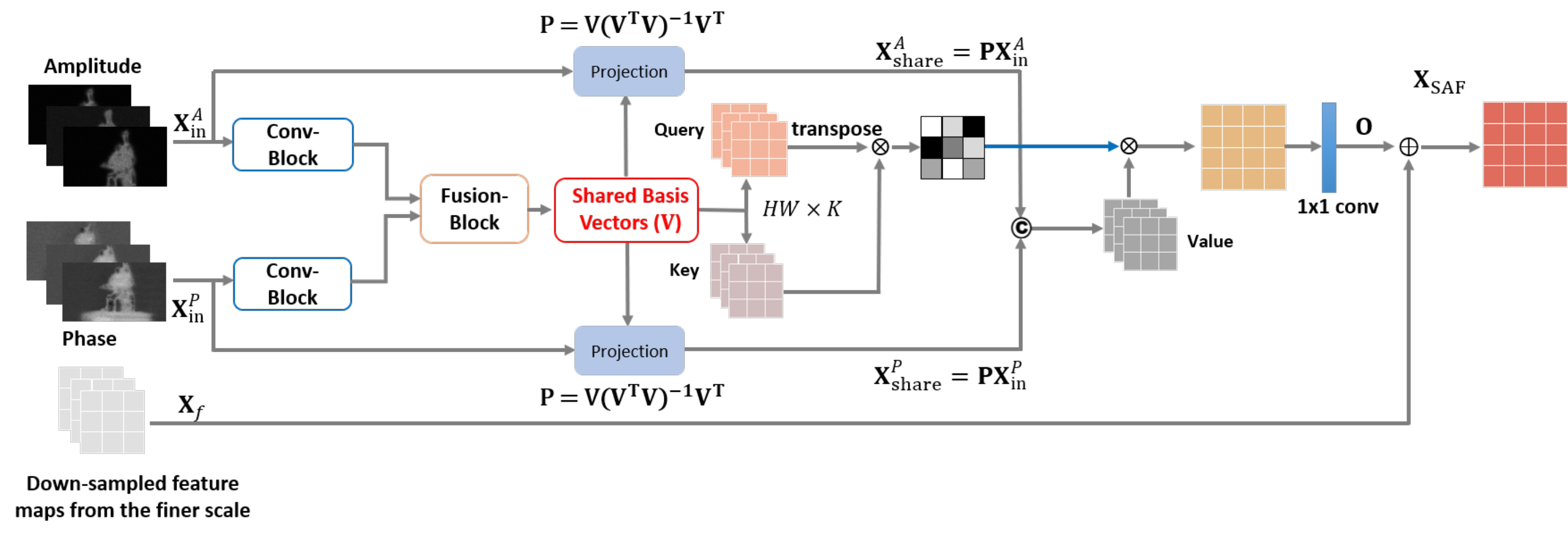}
% \vspace{-0.15in}
\caption{Block diagram of Subspace-Attention-guided Fusion Module (SAFM). SAFM first projects the different-band amplitude and phase features into a common latent subspace and then finds the wide-range dependencies among the projected features via self-attention guidance. As a result, attention-aware features are fused on common ground.} 
\label{fig:s3af}	
% \vspace{-0.2in}
\end{figure*}

\subsection{Intra-view Feature Fusion of \texttt{SARNet}}
\label{sec:intra-view}
\subsubsection{Subspace-Attention-guided Fusion Module}
\label{sec:S3AF}
How to properly fuse the spectral features of THz amplitude and phase are, however, not trivial, as their characteristics can be significantly different. To address the problem, inspired by \cite{cheng2021nbnet} and \cite{zhang2019self}, we propose SAFM shown in Fig.~\ref{fig:s3af} to fuse multi-spectral relevant features on common ground. 
% This sub-network structure consists of two parts: i) we utilize the subset projection to learn the shared basis features from amplitude and phase, and also extract the non-local information for restoration task. ii) we integrate the self-attention module (SA) to learn the long-range information from the shared basis.

Let $\X^{A}_{\mathrm{in}}$, $\X^{P}_{\mathrm{in}} \in \mathbb{R}^{H \times W \times 3}$ denote the spectral bands of the THz amplitude and phase, respectively. The Conv-block $f_C(\cdot)$  extracts two intermediate feature maps $f_C(\X^{A}_{\mathrm{in}})$, $f_C(\X^{P}_{\mathrm{in}}) \in \mathbb{R}^{H \times W \times C_1}$ from $\X^{A}_{\mathrm{in}}$ and $\X^{P}_{\mathrm{in}}$, respectively. As a result, we then derive the $K$ shared basis vectors $\mathbf{V}=[\mathbf{v}_1, \mathbf{v}_2, ..., \mathbf{v}_K]$ from $f_C(\X^{A}_{\mathrm{in}})$ and $f_C(\X^{P}_{\mathrm{in}})$, where $\mathbf{V} \in \mathbb{R}^{N \times K}$, $N=HW$ denotes the dimension of each basis vector, and $K$ is the rank of the shared subspace. The basis set of the shared common subspace is expressed as

% \vspace{-0.1in}
\begin{equation}
    \mathbf{V} = f_F(f_C(\X^{A}_{\mathrm{in}}), f_C(\X^{P}_{\mathrm{in}})),
\label{eq:basis}
\end{equation}
where we first concatenate the two feature maps in the channel dimension and then feed the concatenated feature into the fusion block $f_F(\cdot)$. The structure of the fusion block is the same as that of the Conv-block with $K$ output channels as indicated in the red block in Fig.~\ref{fig:s3af}. The weights of the fusion block are learned in the end-to-end training stage. The shared latent subspace 
learning mainly serves two purposes: 1) learning common latent representations between the THz amplitude and phase bands, and 2) learning the subspace projection matrix to project the amplitude and phase features into the shared subspace such that they can be analyzed on a common ground. These both help identify wide-range dependencies of amplitude and phase features for feature fusion.

To find wide-range dependencies between the amplitude and phase features on common ground, we utilize the orthogonal projection matrix $\mathbf{V}$  in \eqref{eq:basis} to estimate the self-attentions in the shared feature subspace as

\begin{equation}
    \beta_{j,i} = \frac{\exp(s_{ij})}{\sum_{i=1}^{N}\exp(s_{ij})}~, ~s_{ij}= \mathbf{v}_i^{T} \mathbf{v}_j
\label{eq:correlation}
\end{equation}
where $\beta_{j,i}$ represents the model attention in the $i$-th location of the $j$-th region. 

The orthogonal projection matrix $\mathbf{P}$ is derived from the subspace basis $\mathbf{V}$ as follows \cite{meyer10matrix}:

% \vspace{-0.10in}
\begin{equation}
    \mathbf{P} = \mathbf{V}(\mathbf{V}^T\mathbf{V})^{-1} \mathbf{V}^T
\label{eq:orth}
\end{equation}
where $(\mathbf{V}^T\mathbf{V})^{-1}$ is the normalization term to make the basis vectors orthogonal to each other during the basis generation process.

As a result, the output of the self-attention mechanism becomes
\begin{equation}
    \mathbf{o}_{j} = \left( \sum_{i=1}^{N} \beta_{j,i} \mathbf{s}_{i}\right),~ ~\mathbf{s}_{i}=\mathrm{Concate}(\mathbf{P} \X^{A}_{\mathrm{in}}, \mathbf{P} \X^{P}_{\mathrm{in}})
\label{eq:self}
\end{equation}
where the key of $\mathbf{s}_{i}$ $\in$~$\mathbb{R}^{HW \times 6}$ is obtained by concatenating the two feature maps  $\mathbf{PX}^{A}_{\mathrm{in}}$ and  $\mathbf{PX}^{P}_{\mathrm{in}}$ projected by orthogonal projection matrix $\mathbf{P} \in \mathbb{R}^{HW \times HW}$, and $\mathbf{X}^{A}_{\mathrm{in}}$ and $\mathbf{X}^{P}_{\mathrm{in}}$ are reshaped to $HW \times 3$. Since the operations are purely linear with some proper reshaping, they are differentiable.

% Finally, we further multiply the output of the self-attention between two components by a scale parameter and add the down-sampling feature maps from finer scale as:
Finally, we further combine cross-scale features of the self-attention output by adding the down-sampled feature map $\mathbf{X}_f$ from the finer scale as
\begin{equation}
    \mathbf{Y}^{\mathrm{SAF}}_{\mathrm{out}}=f_s(\mathbf{o}) + \mathbf{X}_f
\label{eq:s3af_r}
\end{equation}
where $f_s$ is a 1×1 convolution to keep the channel number consistent with $\mathbf{X}_f$.
% where $\gamma$ is a learnable scalar and it is initialized as 0 from \cite{zhang2019self}.

\begin{figure*}[t]
\centering
\includegraphics[width=0.95\textwidth]{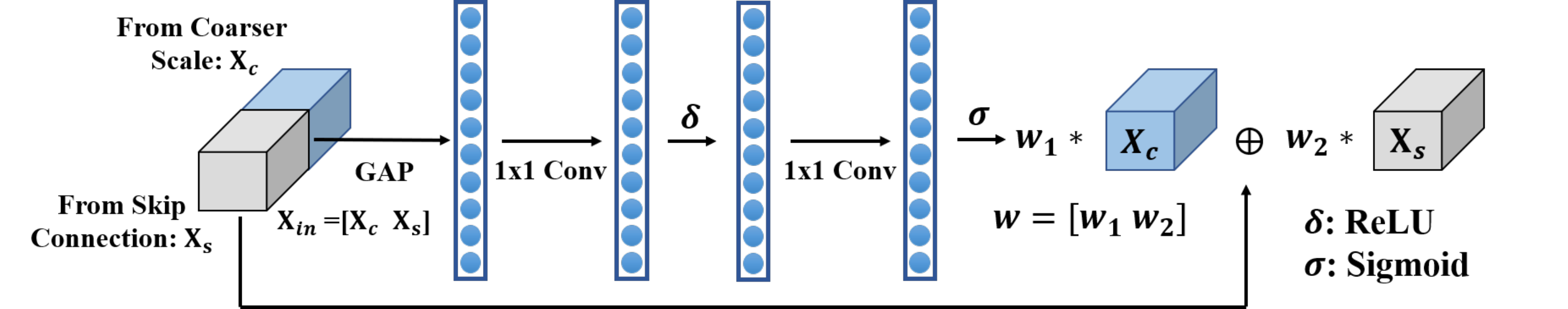}
% \vspace{-0.15in}
\caption{Block diagram of Channel Attention Module (CAM).} 
\label{fig:cam}	
% \vspace{-0.2in}
\end{figure*}

\subsubsection{Channel Attention Module}
\label{sec:CA}

To fuse multi-scale features from different spectral bands in the channel dimension, we incorporate the efficient channel attention mechanism proposed in \cite{qin2020ffa} in the decoder path of \texttt{SARNet} as shown in Fig.~\ref{fig:cam}. In each decoding branch, the original U-Net directly concatenates the up-sampled feature from the coarser scale with the feature from the corresponding encoding branch via the skip-connection shortcut, and then  fuses the intermediate features from different layers by convolutions. This, however, leads to poor image restoration performances in local regions such as incorrect object thickness or details. To address this problem, we propose a channel attention module (CAM) that adopts full channel attention in the dimensionality reduction operation by concatenating two channel attention groups. CAM  first performs global average pooling to extract the global spatial information in each channel:
% \vspace{-0.1in}
\begin{align}
    G_t = \frac{1}{H \times W}\sum_{i=1}^{H}\sum_{j=1}^{W} X_{t}(i,j)
\label{eq:avg}
\end{align}
where $X_{t}(i,j)$ denotes the $t$-th channel of $X_{t}$ at position $(i,j)$ obtained by concatenating the up-sampled feature map $\mathbf{X}_c$ of the coarser-scale and the skip-connection feature map $\mathbf{X}_s$. The shape of $G$ is from $C \times H \times W$ to  $C \times 1 \times 1$. 

% We directly feed the result to perform 1-D convolution  with a kernel size of $M$ to obtain
% \vspace{-0.2in}
% \begin{align}
%     \textbf{w} = \sigma \left(\texttt{C1D}_M(G) \right),
% \label{eq:channel_comp}
% \end{align}
% where $\texttt{C1D}$ represents the 1D convolution, which only involves $M$ parameters. 

We directly feed the result through two stacks of  $1\times 1$ convolutions, sigmoid, and ReLU activation function as:

% \vspace{-0.2in}
\begin{align}
    \mathbf{w} = \sigma \left( \mathrm{Conv}_{1\times1}\left( \delta \left( \mathrm{Conv}_{1\times1}(G) \right) \right) \right),
\label{eq:channel_comp}
\end{align}
where $\mathrm{Conv}_{1\times1}(\cdot)$ denotes a $1\times 1$  convolution, $\sigma$ is the sigmoid function, and $\delta$ is the ReLU function.
In order to better restore a local region, we divide the weights $\textbf{w}$ of different channels into two groups $\mathbf{w}=[\mathbf{w}_1, \mathbf{w}_2]$ corresponding to two different sets of input feature maps, respectively. Finally, we element-wise multiply the input $X_c$ and $X_s$ of the weights $\mathbf{w}$ and add these two group features.
% \begin{align}
%     X_{out} = w_1 X_c +  w_2 X_s
% \label{eq:ass_oper}
% \end{align}
%  where $X_{out}$ is the result after channel attention fusion, $X_c$ and $X_s$ are input feature maps from the up-sampling features in coarser scale in decoder path and encoder features form skip connection from encoder path. 
%  $\X^{A}_{\mathrm{in}}$
\begin{figure*}[!t]
\centering
\includegraphics[width=0.95\textwidth]{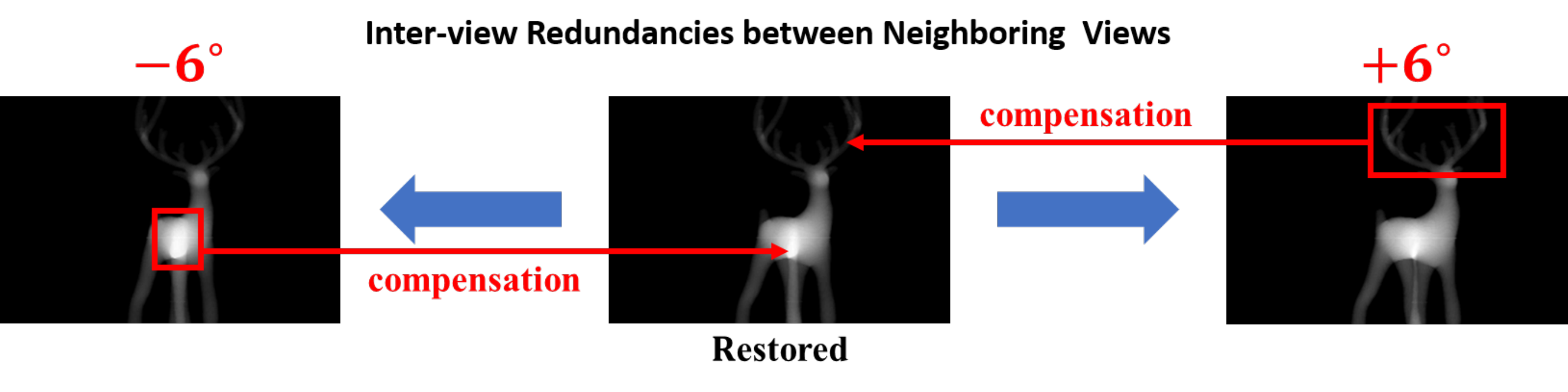}
% \vspace{-0.15in}
\caption{Illustration of inter-view redundancies between neighboring views, where the redundancies existing in the two neighboring views offer useful clues for restoring the center view.} 
\label{fig:inter-view}	
% \vspace{-0.2in}
\end{figure*}

% \subsection{Inter-view Feature Fusion of $\texttt{SARNet}_\texttt{MV}$}
\subsection{Inter-view Feature Fusion of \SMV}
After restoring individual views of an object with \texttt{SARNet}, we then perform multi-view feature fusion between neighboring views to further boost restoration performance. As shown in Fig.~\ref{fig:inter-view}, besides intra-view multi-spectral features, the inter-view redundancies between neighboring views can also provide informative clues for restoring corrupted views. To leverage the inter-view information, as shown in Fig.~\ref{fig:unet}, for the $t$-th corrupted view, we fuse its post-restoration feature tensor with those of its two closest views (\ie the ($t-1$)-th and ($t+1$)-th views with a sampling step-size of $6^\circ$), all restored by the same \texttt{SARNet} model. To achieve inter-view feature fusion, we first concatenate the \texttt{SARNet}-restored feature tensors of three neighboring views centered at the $t$-th view, $\X^{(t-1)}_\mathrm{in}$,  $\X^{(t)}_\mathrm{in}$, and  $\X^{(t+1)}_\mathrm{in}$, as follows:

\begin{align}
    \Hat{\X}^{(t)}_\mathrm{conc} =  \mathrm{Concate}(\X^{(t-1)}_\mathrm{SAF}, \X^{(t)}_\mathrm{SAF}, \X^{(t+1)}_\mathrm{SAF}),
\label{eq:three_cat}
\end{align}
where $\mathrm{Concate}(\cdot)$ denotes the concatenation operation and 
\begin{align}
    \mathbf{X}^{(t)}_{\mathrm{SAF}} = \texttt{SARNet} \left( \mathrm{Conv}_{3\times3}(\mathbf{X}^{(t)}_{\mathrm{in}}), \mathbb{W}(\X^{(t)}_{\mathrm{in}}) \right), 
\label{eq:three_sarnet2}
\end{align}
where $\texttt{SARNet}(\cdot)$ denotes the restoration model, $\mathrm{Conv_{3\times3}(\cdot)}$ denotes a $3\times 3$  convolution, and  $\mathbb{W}(\X^{(t)}_{\mathrm{in}})$ denoted the set of 24 amplitude and phase spectral bands of $\X^{(t)}_{\mathrm{in}}$ selected based on physics guidance.

The concatenated three-view feature tensor is then fused via the feature fusion  block involving a CAM and a Cov-Block as follows:

\begin{align}
    \X^{(t)}_{\mathrm{MVF}} =  \mathrm{Conv}_{3\times3} \left( \mathrm{CAM}\left( \Hat{\X}^{(t)}_\mathrm{conc} \right) \right),
\label{eq:three_fusion}
\end{align}
where $\X^{(t)}_{\mathrm{MVF}}$ denotes the multi-view fused version of the $t$-th view, and $\mathrm{CAM(\cdot)}$ is the channel attention module mentioned above. 

Finally, as illustrated in Fig.~ we feed the fused three-view feature tensor $\mathbf{X}^{t}_{\mathrm{MVF}}$ into the finest-scale branch of  \texttt{SARNet} along with the 24 amplitude and phase bands (\ie the water absorption profile)  associated with $\X^{t}_\mathrm{in}$ to obtain the final restoration result $\mathbf{X}^{t}_{\mathrm{rec}}$ as

\begin{align}
    \mathbf{X}^{(t)}_{\mathrm{rec}} = \texttt{SARNet} \left( \mathbf{X}^{(t)}_{\mathrm{MVF}}, \mathbb{W}(\X^{(t)}_{\mathrm{in}}) \right). 
\label{eq:three_sarnet3}
\end{align}

%where the \texttt{SARNet} model $\mathbf{F}_{\mathrm{Stage2}}(\cdot)$ is as same as the $\mathbf{F}_{\mathrm{Stage1}}(\cdot)$ from first stage. 

\subsection{Loss Function for THz Image Restoration}
\label{sec:loss}
To effectively train \texttt{SARNet}, we employ the following mean squared error (MSE) loss function to measure the dissimilarity between the restored image $\mathbf{X}_{\mathrm{rec}}$ and its ground-truth $\mathbf{X}_{\mathrm{GT}}$:
	\begin{align}
	\mathcal{L}_{\mathrm{MSE}}(\mathbf{X}_{\mathrm{GT}},\mathbf{X}_{\mathrm{rec}}) = &\frac{1}{HW}\sum_{i=1}^{H}\sum_{j=1}^{W}(\mathbf{X}_{\mathrm{GT}}(i,j)-\mathbf{X}_{\mathrm{rec}}(i,j))^2,
	\label{eq:loss}
	\end{align}
where  $H$ and $W$ are the height and width of the image.

 \subsection{3D Tomographic Reconstruction}
 \label{sec:3d}
  The 3D tomography of an object is reconstructed from the 60 2-D restored views of the object scanned from different angles.  To reconstruct a 3-D tomography from the 60 2-D views, we directly utilize the inverse Radon transform to obtain the 3-D tomography, using methods like filtered back-projection (FPB) \cite{kak2001algorithms} or the simultaneous algebraic reconstruction technique (SART) \cite{recur2011investigation}.

\section{Experimental Results} 
\label{sec:experimentS}

We conduct experiments to evaluate the effectiveness of \texttt{SARNet} against existing state-of-the-art restoration methods. We first present our THz-TDS system and measurement. Then, the details of the THz dataset and experiment settings. Finally, we evaluate the performances of \texttt{SARNet} and the competing methods on THz image restoration and tomographic reconstruction.

\subsection{Proposed ASOPS THz-TDS System}
\label{sec:thz-tds}
Our in-house THz measurement system is an asynchronous optical sampling THz time-domain spectroscopy system (ASOPS THz-TDS), which is composed of two asynchronous femtosecond lasers whose central wavelength are located at 1550 nm with tens of mW level, a pair of THz photoconductive antenna (THz PCA) source and detector, a linear and rotation motorized stage, four plane-convex THz lens with 50 mm focal length, a transimpedance amplifier (TIA), and a unit of data acquisition (DAQ) and processing~\cite{janke2005asynchronous}. 
\textcolor{blue}{}
The repetition rates of the two asynchronous femtosecond lasers are 100 MHz and 100 MHz + 200 Hz, respectively. 
The sampling rate of DAQ is 20 MHz. With the configuration above, our ASOPS THz-TDS system delivers 0.1 ps temporal resolution and a THz frequency bandwidth of 5 THz. 
Additionally, our ASOPS THz-TDS system provides THz pulse signals with 41.7 dB dynamic range from 0.3 THz to 3 THz and 516 femtoseconds at full width at half maximum (FWHM). 
However, under the configuration above, the number of sampling points for a trace is approximately 100 K, consuming an extremely large transmission bandwidth. 
To address this limitation, only the 100-ps segment of the THz pulse signal is extracted. With the extracted segment of 100 ps, the frequency resolution is 10 GHz. 
Additionally, considering the minimum THz beam diameter of about 1.25 mm and the diffraction limitation, our THz system can provide spatial resolution in the scale of sub-millimeters.\\

\begin{figure}[t]
\centering
\includegraphics[width=1\textwidth]{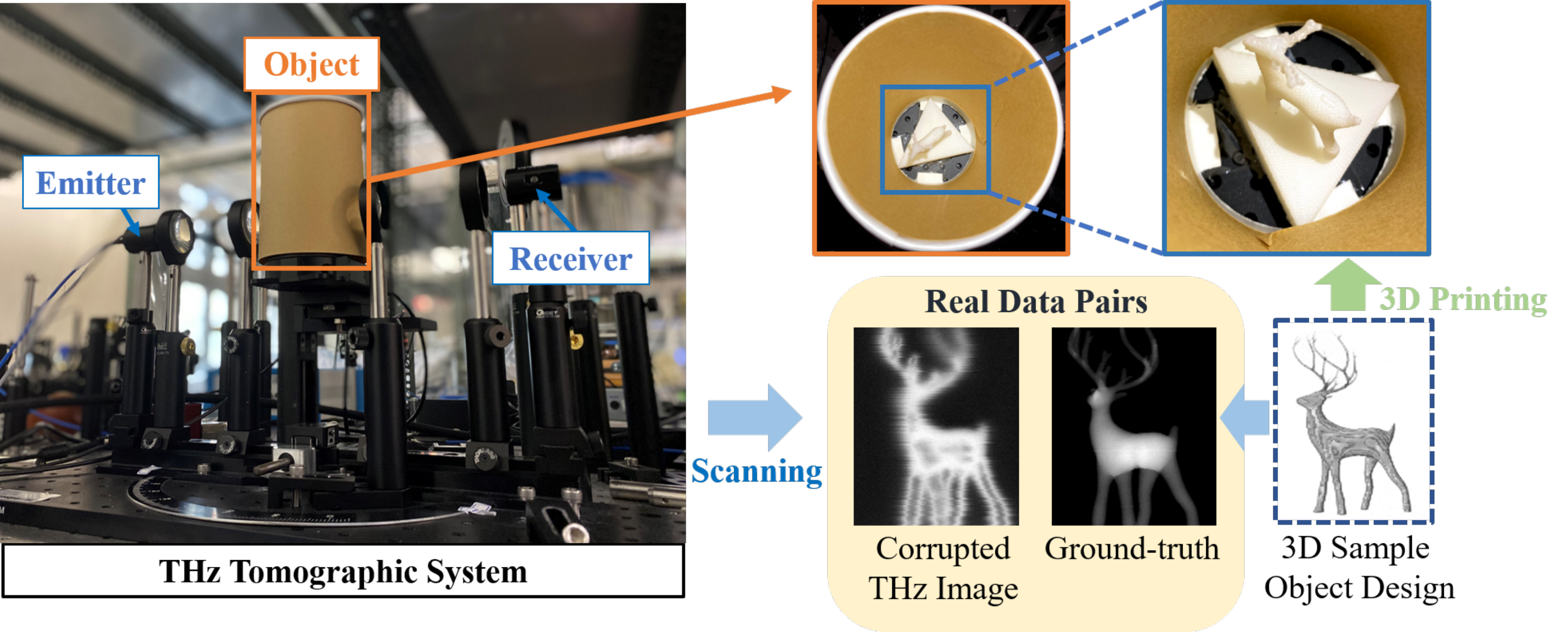}
% \vspace{-0.3in}
\caption{Illustration of THz data collection with our in-house THz-TDS tomographic imaging system.} 
\label{fig:design}	
% \vspace{-0.3 in}
\end{figure}

\subsection{Properties of THz  Measurements}
\label{sec:properties}
\begin{figure}
    \centering
    \includegraphics[scale=0.4]{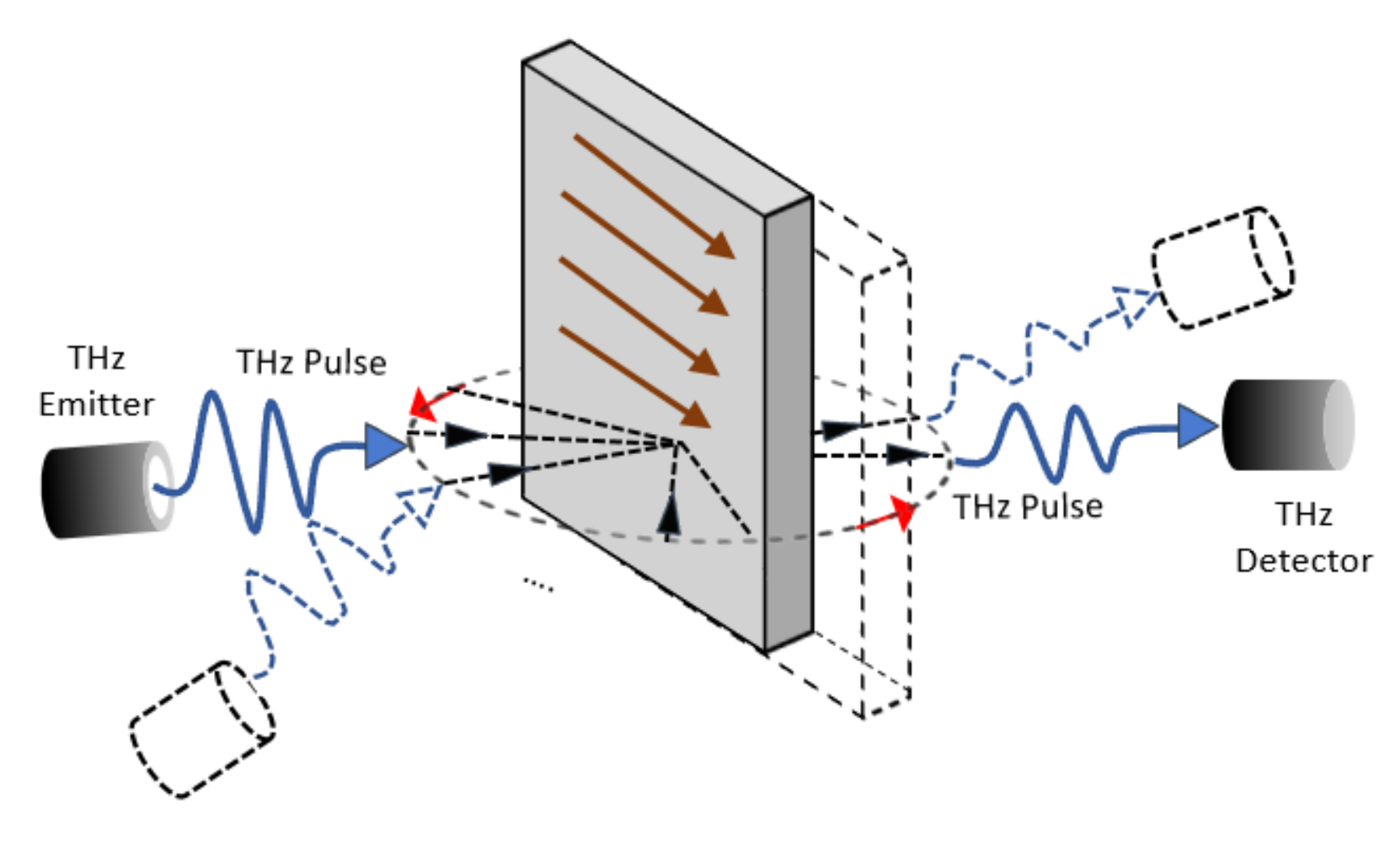}
    \caption{The THz image formation system based on the raster scanning approach.}
    \label{fig:image formation}
\end{figure}

To retrieve the temporal-spatial-spectral information of each object voxel, our THz imaging experiment setup is based on a THz-TDS system as shown in Fig.~\ref{fig:design}. 
To demonstrate the THz penetrating capability, the measured object is first covered by a paper shield, which is highly transparent to THz but opaque in visible light.
The covered object (\eg a 3D printed deer covered by a paper shield) is placed on the rotation stage in the THz path between the THz source and detector of the THz-TDS system and is scanned by a raster scanning approach in 60 projection angles, as shown in Fig. \ref{fig:image formation}. %

During measuring, the THz-TDS system profiles each voxel's THz temporal signal with 0.1 ps temporal resolution, whose amplitude corresponds to the strength of the THz electric field. 
With this scanning approach, a cube object of size 2 cm $\times$ 2 cm $\times$ 2 cm consumes about 1 min for scanning a projected 2D image; thus, the cube will take about an hour for the 60 projection angles.
Additionally, due to the limitation of the linear motorized stage, our measuring system can support an object size of about 6 cm at maximum.
With our THz imaging experiment setup, the THz beam diameter varies with the THz propagation direction. As a consequence, the point-spread function of our system will vary with the geometry and location of the object.
Therefore, the 2D projected images of the thickness-varying object could suffer from different levels of blurring effects in different pixels.
%
%Besides, considering the interaction between water and THz beams, we extract the amplitude and image images at 12 frequencies based on the water-absorption lines: 0.380, 0.448, 0.557, 0.621, 0.916, 0.970, 0.988, 1.097, 1.113, 1.163, 1.208, and 1.229 THz.

\begin{figure}[t]
\centering
\includegraphics[width=1\textwidth]{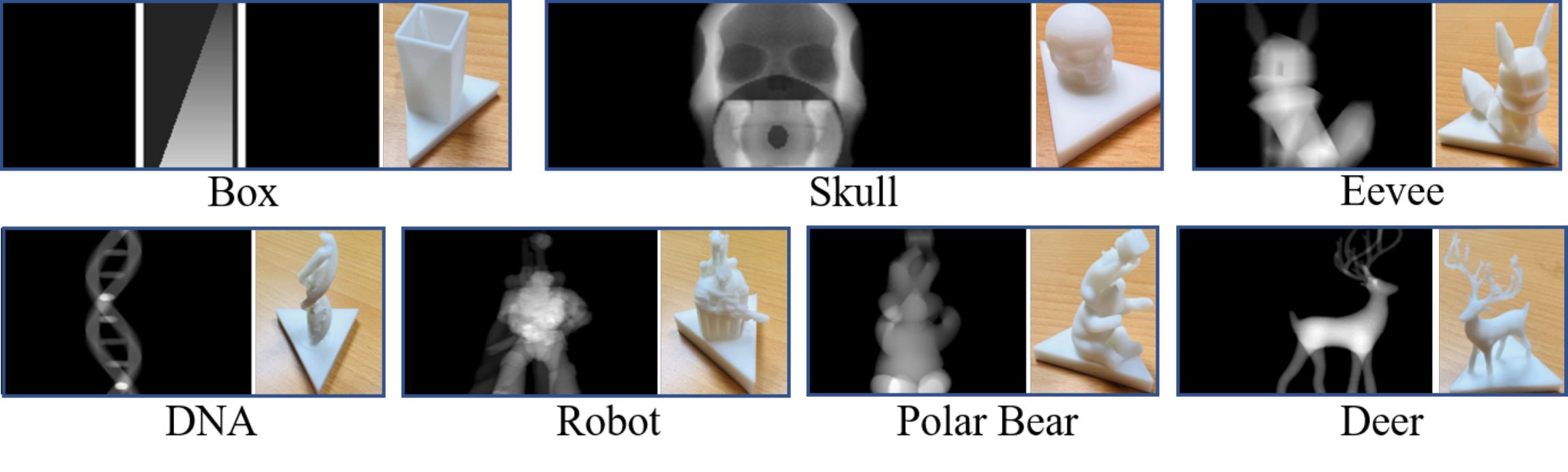}
% \vspace{-0.2in}
\caption{Illustration of the ground-truths and photos of the seven 3D-printed HIPS objects used in our experiments. The left image of each object illustrates the ground-truth of one projection view and the right shows the photo of the HIPS object.} 
\label{fig:data}	
% \vspace{-0.2in}
\end{figure}

\subsection{THz-TDS Image Dataset}
As shown in Fig.~\ref{fig:design}, we prepare the  sample objects by a Printech 3D printer, and use the material of high impact polystyrene (HIPS) for 3D-printing the objects.

The HIPS material is chosen since it can be used to quickly fabricate target objects by cost-efficient 3D printers, which can help evaluate a wide range of object geometries.
Additionally, the low absorption nature of HIPS in the THz range can prevent severe SNR degradation of detected THz signals while scanning objects.
We then use our in-house ASOPS THz-TDS system~\cite{janke2005asynchronous} presented in Sec.~\ref{sec:thz-tds} to measure the sample objects. Each sample object is placed on a motorized stage between the source and the receiver. With the help of the motorized stage, raster scans are performed on each object in multiple view angles. In the scanning phase, we scan the objects covering a rotational range of 180 degrees (step-size: 6 degrees between two neighboring views),  a horizontal range of 72mm (step-size: 0.25mm), and a variable vertical range corresponding to the object height (step-size: 0.25mm). In this way, we obtain 30 projections of each object, which are then augmented to 60 projections by horizontal flipping. 
The ground-truths of individual projections are obtained by taking the Radon transform of the 3D digital models defining the 3D object profiles for 3D printing in every view-angle.
In addition to generating from digital models, the ground-truths can also be generated through precise 3D scanners.

We use markers to indicate the center of rotation so that we can align the ground-truths with the measured THz data. In this paper, totally seven sample objects are printed, measured, and aligned for evaluation. 

% \vspace{-0.05in}
\subsection{Data Processing and Augmentation}
In our experiments, we train the proposed multi-view $\texttt{SARNet}_{\texttt{MV}}$ model using the 2D THz images collected from our THz imaging system shown in Fig.~\ref{fig:band}. Fig.~\ref{fig:data} illustrates the photos of seven example objects along with their 2D ground-truths at certain projection angles. Each object consists of 60 projections and there are 420 2D THz images in total. In order to thoroughly evaluate the capacity of $\texttt{SARNet}_{\texttt{MV}}$, we adopt the leave-one-out strategy: using the data of 6 objects (\ie 360 training images) as the training set, and that of the remaining object as the testing set. Due to the limited space, we only present part of the results in this section, and the complete results in the supplementary material and our project site\footnote{Project site: \href{https://github.com/wtnthu/THz\_Tomography}{https://github.com/wtnthu/THz\_Tomography}}. The THz-TDS image dataset can be found in the dataset site\footnote{Dataset site: \href{https://github.com/wtnthu/THz\_Data}{https://github.com/wtnthu/THz\_Data}}. We will release our source code after the paper is accepted.

We also perform typical data augmentations to enrich the training set, including random rotating and flipping. Finally, the images are randomly cropped to $128 \times 128$ patches.

% We also perform typical data augmentations to enrich the training set, involving random transformations on brightness, contrast, and saturation, image scaling with a scaling factor ranging in $[0.5, 1.5]$, and geometric transformations including random flipping horizontally and vertically. Finally, the images are randomly cropped to $128 \times 128$ patches.

% \vspace{-0.2in}
\subsection{Experiment Settings}
We initialize $\texttt{SARNet}_{\texttt{MV}}$ following the initialization method in \cite{he2015delving}, and train it using the Adam optimizer with $\beta_1 = 0.9$ and $\beta_2 = 0.999$. We set the initial learning rate to $10^{-4}$ and then decay the learning rate by $0.1$ every $300$ epochs. \texttt{SARNet} converges after $1,000$ epochs.  For a fair comparison with the competing methods, we adopt their publicly released codes. All experiments were performed in a Python environment and Pytorch package
running on a PC with Intel Core i7-10700 CPU 2.9 GHz and an Nvidia Titan 2080 Ti GPU.

\subsection{Quantitative and Qualitative Evaluations}
% \vspace{-0.1in}

To the best of our knowledge, there is no method  specially designed for restoring THz images besides $\texttt{Time-max}$~\cite{hung2019terahertz}. Thus, we compare our $\texttt{SARNet}$ and $\texttt{SARNet}_{\texttt{MV}}$ models against several representative CNN-based image restoration models, including $\texttt{DnCNN}$~\cite{zhang2017beyond}, $\texttt{RED}$~\cite{mao2016image}, $\texttt{NBNet}$~\cite{cheng2021nbnet}, and $\texttt{U-Net}$~\cite{ronneberger2015u}. We use the \texttt{time-max} images as the inputs and their corresponding round-truths as the target outputs to train these CNN-based image restoration models. Note that, these CNN-based restoration models do not utilize the prominent spectral information based on water absorption lines for restoring the THz \texttt{time-max} images.
% Moreover, we also compare two variants of U-Net \cite{ronneberger2015u}: baseline U-Net ($\texttt{U-Net}_\texttt{base}$) and multi-spectral U-Net ($\texttt{U-Net}_\texttt{MS}$).
% $\texttt{U-Net}_\texttt{base}$  extracts image features in five different scales following the original setting in U-Net \cite{ronneberger2015u}, whereas $\texttt{U-Net}_\texttt{MS}$ incorporates multi-spectral features by concatenating the features of \texttt{Time-max} image with additional 12 THz bands for amplitude as the input (i.e., $12+1$ channels) of the finest scale of U-Net. %In addition, we compare the other variants of Unet in next sub-section which incorporates Time-max with phase and simultaneous amplitude and phase as inputs, respectively. 
For quantitative quality assessment, we adopt two quality metrics for assessing the visual qualities of 2D view restoration and 3D tomographic reconstruction, respectively. and the reconstruction quality. The first metric is the Peak Signal-to-Noise Ratio (PSNR) for measuring the discrepancy between restored 2D views and their ground-truth as shown in Table\,\ref{table:PSNR}. The second is the Mean-Square Error (MSE) between the cross-sections of a reconstructed 3D tomography and the corresponding ground-truths for assessing the 3D reconstruction accuracy as compared in Table\,\ref{table:3DMSE}. To further evaluate the 3D reconstruction accuracy of various models, as shown in Table\,\ref{table:3DIOU}, we also  compare the average Intersection over Union (IoU), F-Score, and Chamfer distance performances by converting reconstructed 3D volumes into point-clouds \cite{xie2020pix2vox}.
%\red{Our main task is to do 3D tomography, and the correct thickness of the 2D image will directly affect the 3D tomography. To estimate the thickness of the tomography, we further evaluate of each cross-section of tomography results for all 7 objects by using the mean square error (MSE) in Table.\,\ref{table:3DMSE}.}
% Additionally, our main task is to do 3D tomography, and the correct thickness of the 2D image will directly affect the 3D tomography.

Table\,\ref{table:PSNR} shows that our  $\texttt{SARNet}$ and $\texttt{SARNet}_{\texttt{MV}}$ both significantly outperform the competing methods on all the seven sample objects in PSNR. 
Specifically, $\texttt{SARNet}_{\texttt{MV}}$ outperforms $\texttt{Time-max}$~\cite{hung2019terahertz}, $\texttt{U-Net}$~\cite{ronneberger2015u}, and $\texttt{NBNet}$~\cite{cheng2021nbnet} by 11.41 dB, 2.79 dB, and 2.23 dB, respectively, in average PSNR. In particular, even based on a simpler backbone $\texttt{U-Net}$, thanks to the good exploration of physics guidance, our proposed models significantly outperform  the state-of-the-art restoration model $\texttt{NBNet}$ especially on challenging objects like \textbf{Box} and \textbf{Skull}. With the aid of inter-view redundancies, multi-view $\texttt{SARNet}_{\texttt{MV}}$ stably outperforms single-view $\texttt{SARNet}$ and achieves notable 0.56 dB and 0.57 dB PSNR gains on \textbf{Box} and \textbf{Skull}. Similarly, in terms of 3D reconstruction accuracy, Table\,\ref{table:3DMSE} demonstrates that our models both stably achieve significantly lower average MSE of tomographic reconstruction than the competing methods on all seven objects. As for 3D shape reconstruction accuracy, Table \ref{table:3DIOU} demonstrates that our models stably achieve significantly higher performances, in terms of average IoU, F-Score, and Chamfer distance of tomographic reconstruction, than the competing methods for all the seven objects.

% Specifically, our S3AF-Unet outperforms Time-max,  baseline U-Net (U-Net$_{\text{base}}$), and the multi-spectral U-Net (U-Net$_{\text{MS}}$) by 10.42 dB, 3.00 dB, and 0.38 dB in PSNR for \textbf{Deer}, and 14.27 dB, 2.19 dB, and 1.29 dB for \textbf{DNA}, respectively.
\begin{table}[!]
\begin{center}
\begin{minipage}{\textwidth}
\caption{Quantitative comparison of PSNR between restored 2D views and their ground-truth  with different methods on \textbf{Deer}, \textbf{DNA}, \textbf{Box}, \textbf{Eevee}, \textbf{Polarbear}, \textbf{Robot}, and \textbf{Skull}.  ($\uparrow$: higher is better)}
\label{table:PSNR}
\begin{tabular*}{\textwidth}{@{\extracolsep{\fill}}lccccccc@{\extracolsep{\fill}}}
\toprule%
& \multicolumn{7}{@{}c@{}}{PSNR (dB) $\uparrow$}  \\\cmidrule{2-8}%
Method   &Deer   &DNA    &Box     &Eevee    &Polarbear    &Robot   &Skull \\ \hline 
\midrule
				    $\texttt{Time-max}$                             &12.42  &12.07  &11.97   &11.20    &11.21        &11.37   &10.69 \\ \hline
					$\texttt{DnCNN-S}$ \cite{zhang2017beyond}       &19.94  &23.95  &19.13   &19.69    &19.44        &19.72   &17.33  \\ \hline
					$\texttt{RED}$ \cite{mao2016image}              &19.30	 &24.17  &20.18	  &19.97    &19.17        &19.76   &16.28  \\ \hline
					$\texttt{NBNet}$ \cite{cheng2021nbnet}          &20.24  &25.10  &20.21   &19.84    &20.12        &20.01      &19.69     \\ \hline
					$\texttt{U-Net}$ \cite{ronneberger2015u}    &19.84	 &24.15  &19.77	  &19.95    &19.09        &18.80   &17.49  \\ \hline
				% 	$\texttt{U-Net}_\texttt{MS}$                              &22.46	 &25.05  &20.81	  &20.34    &19.86        &20.64   &19.43  \\ \hline 
					\texttt{SARNet} (Ours)         &\underline{22.98}	 &\underline{26.05}  &\underline{22.67}  &\underline{20.87}   &\underline{21.42}   &\underline{22.66}    &\underline{22.48} \\
					\hline
					$\texttt{SARNet}_{\texttt{MV}}$  (Ours) &\textbf{23.17}	 &\textbf{26.19}  &\textbf{23.23}  &\textbf{20.97}   &\textbf{21.55}   &\textbf{22.68}    &\textbf{23.05} \\ \hline
\botrule
\end{tabular*}
\end{minipage}
\end{center}
%\label{tab:t1}
\end{table}

\begin{table}[!]
\begin{center}
\begin{minipage}{\textwidth}
\caption{Quantitative comparison of MSE between the cross-sections of a reconstructed 3D tomography and the corresponding ground-truths with different methods on \textbf{Deer}, \textbf{DNA}, \textbf{Box}, \textbf{Eevee}, \textbf{Polarbear}, \textbf{Robot}, and \textbf{Skull}.  ($\downarrow$: lower is better)}
\label{table:3DMSE}
\begin{tabular*}{\textwidth}{@{\extracolsep{\fill}}lccccccc@{\extracolsep{\fill}}}
\toprule%
& \multicolumn{7}{@{}c@{}}{MSE $\downarrow$}  \\\cmidrule{2-8}%
                    Method                                                   &Deer   &DNA    &Box     &Eevee    &Polarbear    &Robot   &Skull \\ \hline 
\midrule
				    $\texttt{Time-max}$                                      &0.301  &0.026  &0.178   &0.169    &0.084        &0.203   &0.225 \\ \hline
					$\texttt{DnCNN-S}$ \cite{zhang2017beyond}                &0.153  &0.162  &0.309   &0.149    &0.056        &0.223   &0.293  \\ \hline
					$\texttt{RED}$ \cite{mao2016image}                       &0.139	 &0.238  &0.300	  &0.179    &0.070        &0.215   &0.324  \\ \hline
					$\texttt{NBNet}$ \cite{cheng2021nbnet}                   &0.240  &0.184  &0.305   &0.134    &0.088        &0.128   &0.138     \\ \hline
					$\texttt{U-Net}$ \cite{ronneberger2015u}   &0.227	 &0.166  &0.266	  &0.157    &0.077        &0.093   &0.319  \\ \hline
				% 	$\texttt{U-Net}_\texttt{MS}$                              &22.46	 &25.05  &20.81	  &20.34    &19.86        &20.64   &19.43  \\ \hline 
					\texttt{SARNet} (Ours)         &\underline{0.107}	 &\underline{0.015}  &\underline{0.041}  &\underline{0.122}   &\underline{0.050}   &\underline{0.065}    &\underline{0.052} \\
					\hline
					$\texttt{SARNet}_{\texttt{MV}}$  (Ours) &\textbf{0.091}	 &\textbf{0.013}  &\textbf{0.030}  &\textbf{0.105}   &\textbf{0.038}   &\textbf{0.059}    &\textbf{0.049} \\ \hline
\botrule
\end{tabular*}
\end{minipage}
\end{center}
%\label{tab:3d}
\end{table}

\begin{figure*}[!hbt]
\centering
	
\caption*{\tiny  \ \ \ \ \ \ $\texttt{Time-max}$\ \ \ \ \ \ \ \ \  $\texttt{DnCNN-S}$\ \ \ \ \ \ \ \ \ \ \  $\texttt{RED}$  \ \ \ \ \ \ \ \ \ \ \ \ \ $\texttt{NBNet}$ \ \ \ \ \ \ \  $\texttt{U-Net}_{\texttt{base}}$\ \ \ \ \ \ \ \ \  $\texttt{SARNet}$\ \ \ \ \ \ \ \ \ \ \  $\texttt{SARNet}_{\texttt{MV}}$ \ \ \ \ \ \ \ \ \ \ \ GT \ \ \ \ \ \ }
\includegraphics[width=1\textwidth]{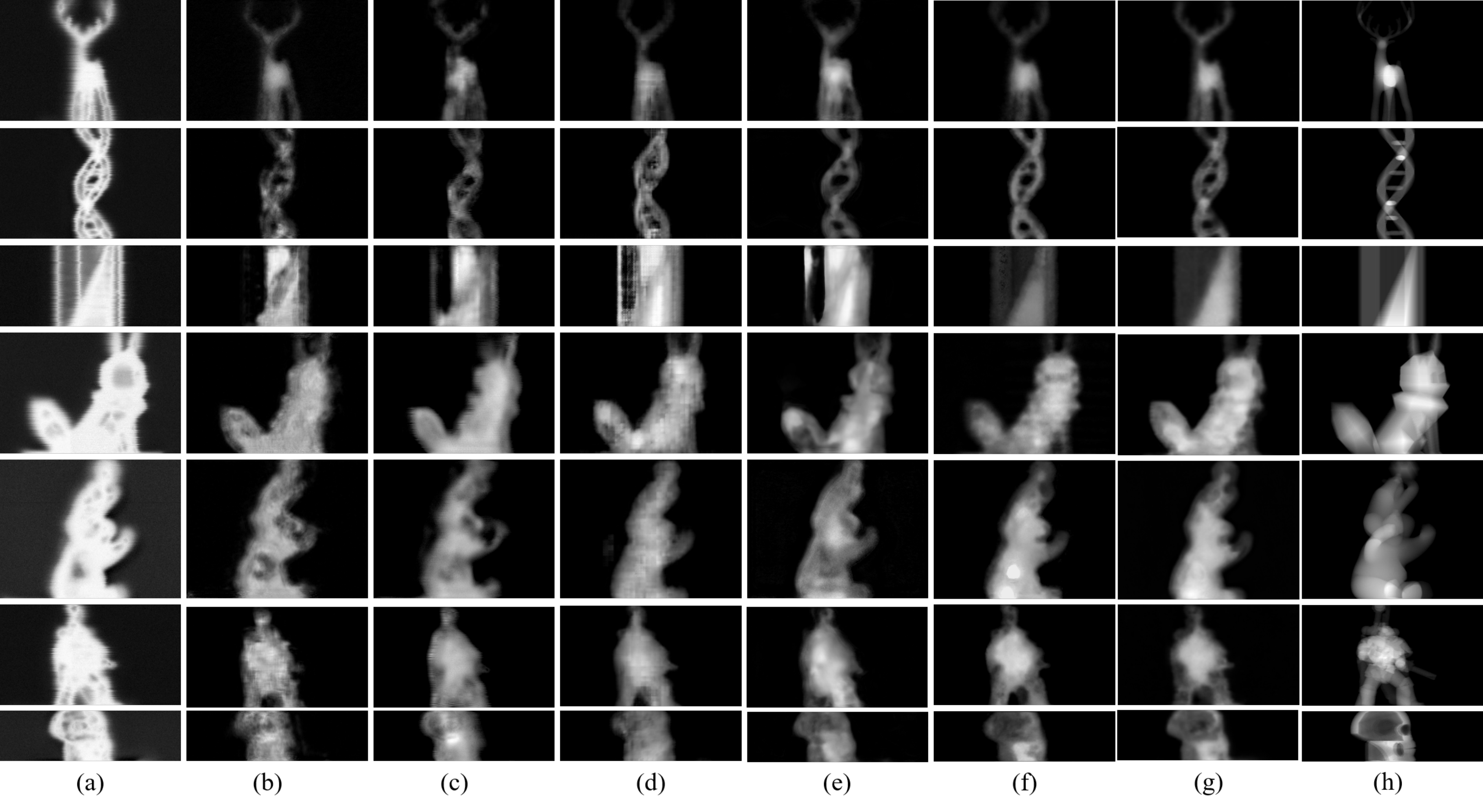}
% \vspace{-0.3in}
\caption{Qualitative comparison of THz image restoration results for \textbf{Deer}, \textbf{DNA}, \textbf{Box}, \textbf{Eevee}, \textbf{Polarbear}, \textbf{Robot}, and \textbf{Skull} from left to right: (a) $\texttt{Time-max}$, (b) $\texttt{DnCNN-S}$ \cite{zhang2017beyond}, (c) $\texttt{RED}$ \cite{mao2016image}, (d) $\texttt{NBNet}$ \cite{cheng2021nbnet}, (e) $\texttt{U-Net}_{\texttt{base}}$ \cite{ronneberger2015u}, (f) $\texttt{SARNet}$, (g) $\texttt{SARNet}_{\texttt{MV}}$, and (h) the ground-truth.} 
\label{fig:vis1}	
% \vspace{-0.25in}
\end{figure*}

For qualitative evaluation, Fig.\,\ref{fig:vis1} illustrates a few restored views for the seven sample objects, demonstrating that $\texttt{SARNet}_{\texttt{MV}}$ can restore objects with much finer and smoother details (\eg the antler and legs of \textbf{Deer}, the base pairs and shapes of \textbf{DNA} double-helix, the depth and shape of \textbf{Box}, the body and gun of \textbf{Robot}, and the correct depth and of \textbf{Skull}),  the faithful thickness of material (\eg the body and legs of \textbf{Deer}, and the correct edge thickness of \textbf{Box}), and fewer artifacts (e.g., holes and broken parts). 

\begin{figure*}[t!]
\centering
\includegraphics[width=1\textwidth]{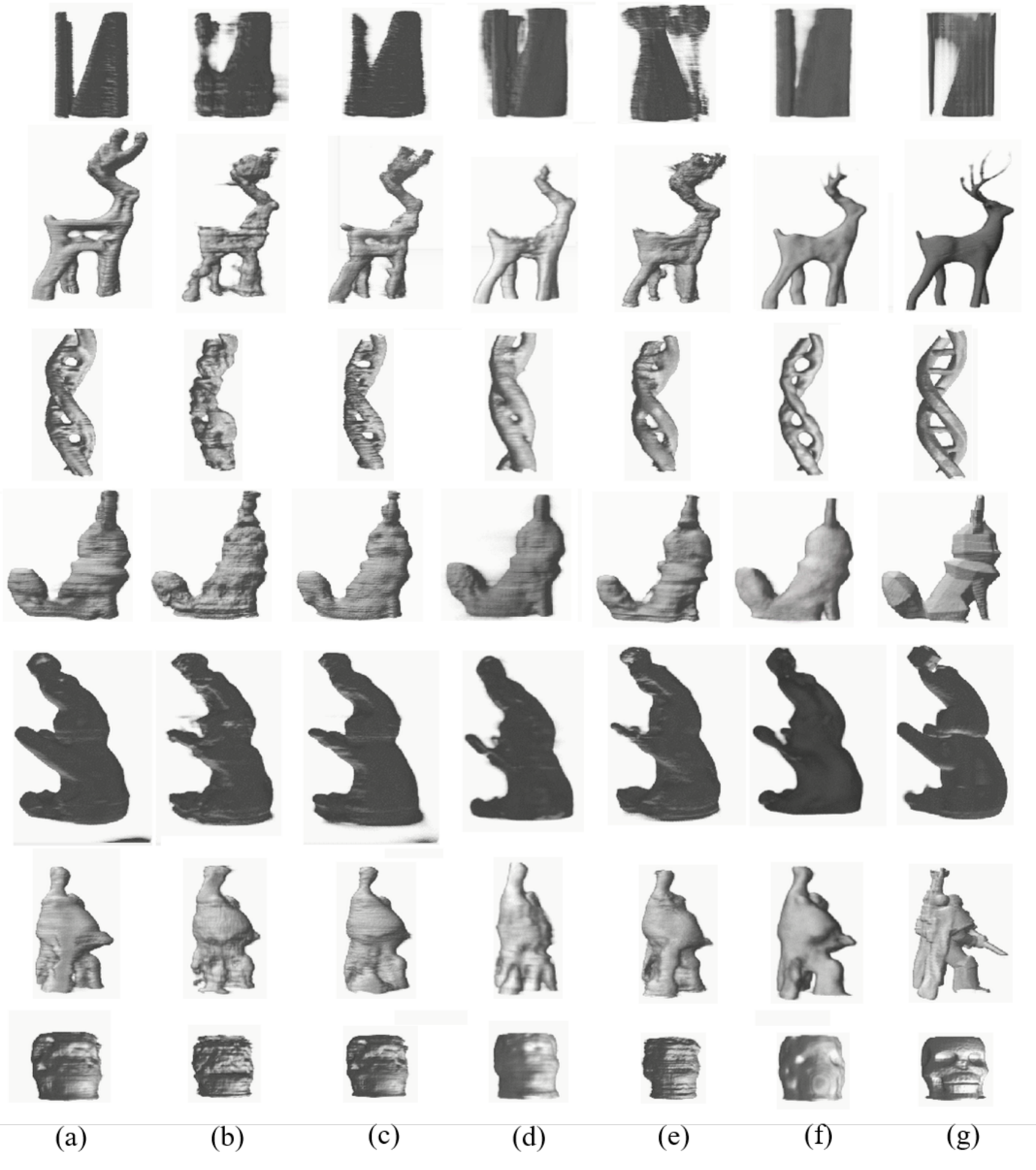}
% \vspace{-0.3in}
\caption{Illustration of 3D tomographic reconstruction results on \textbf{Box}, \textbf{Deer}, \textbf{Dna}, \textbf{Eevee}, \textbf{Polarbear}, \textbf{Robot}, and \textbf{Skull} from left to right: (a) $\texttt{Time-max}$, (b) $\texttt{DnCNN-S}$ \cite{zhang2017beyond}, (c) $\texttt{RED}$ \cite{mao2016image}, (d) $\texttt{NBNet}$ \cite{cheng2021nbnet}, (e) $\texttt{U-Net}_\texttt{base}$ \cite{ronneberger2015u}, (f) $\texttt{SARNet}_{\texttt{MV}}$,  and (g) the ground-truth.} 
\label{fig:3d}	
% \vspace{-0.25in}
\end{figure*}

Our THz tomographic imaging system aims to reconstruct clear and faithful 3D object shapes. 
In our system, the tomography of an object is reconstructed from 60 views of 2D THz images of the object, each being restored by various image restoration models, via the inverse Radon transform. 
The paper shield region is cropped out to mitigate the evaluation bias caused by the simple geometry of the covered paper shield.
Fig.~\ref{fig:3d} illustrates the 3D reconstructions of the seven sample objects, showing that \texttt{Time-max},  $\texttt{U-Net}$ tend to lose important object details such as holes in the deer's body with \texttt{Time-max} and the severely distorted antlers and legs with the three methods. In contrast, our method reconstructs much clearer and more faithful 3D images with finer details, achieving by far the best 3D THz tomography reconstruction quality in the literature. Complete 3D reconstruction results are provided in the supplementary material.

Both the above quantitative and qualitative evaluations confirm a significant performance leap with $\texttt{SARNet}_{\texttt{MV}}$ over the competing methods. Compared with our single-view restoration model (\texttt{SARNet}), the multi-view model $\texttt{SARNet}_{\texttt{MV}}$ can restore finer local details such as the thickness of clear antlers, thinner edge of the box, and the gun in robot's hand. This also means that the inter-view redundancies between neighboring views are helpful in restoring local details, especially since our main task is to do 3D tomography. The correct thickness of the 2D image will directly affect the 3D tomography.

\begin{table}
\begin{center}
\begin{minipage}{\textwidth}
\caption{Quantitative comparison of average IoU, F-Score, and Chamfer distance between the 3D volume of a reconstructed 3D tomography and the corresponding ground-truths with different methods on \textbf{Deer}, \textbf{DNA}, \textbf{Box}, \textbf{Eevee}, \textbf{Polarbear}, \textbf{Robot}, and \textbf{Skull}.  ($\uparrow$: higher is better and $\downarrow$: lower is better)}
\label{table:3DIOU}
\begin{tabular*}{\textwidth}{@{\extracolsep{\fill}}lccccccc@{\extracolsep{\fill}}}
\toprule\midrule%
& \multicolumn{7}{@{}c@{}}{IoU $\uparrow$}  \\\cmidrule{2-8}%
                    Method                                                   &Deer   &DNA    &Box     &Eevee    &Polarbear    &Robot   &Skull \\ \hline 
\midrule
				    $\texttt{Time-max}$                                      &0.247  &0.427  &0.106   &0.323    &0.482        &0.041   &0.385 \\ \hline
					$\texttt{DnCNN-S}$ \cite{zhang2017beyond}                &0.179  &0.136  &0.096   &0.509    &0.353        &0.260   &0.158  \\ \hline
					$\texttt{RED}$ \cite{mao2016image}                       &0.386	 &0.323  &0.257	  &0.359    &0.433        &0.142   &0.175  \\ \hline
					$\texttt{NBNet}$ \cite{cheng2021nbnet}                   &0.255  &0.163  &0.414   &0.633    &0.526        &0.170   &0.419     \\ \hline
					$\texttt{U-Net}$ \cite{ronneberger2015u}   &0.400	 &0.427  &0.117	  &0.423    &0.539        &0.290   &0.286  \\ \hline
				% 	$\texttt{U-Net}_\texttt{MS}$                              &22.46	 &25.05  &20.81	  &20.34    &19.86        &20.64   &19.43  \\ \hline 
					\texttt{SARNet} (Ours)         &0.502	 &0.515  &0.418  &0.702   &0.550   &0.434    &0.407 \\
					\hline
					$\texttt{SARNet}_{\texttt{MV}}$  (Ours) &\textbf{0.538}	 &\textbf{0.567}  &\textbf{0.424}  &\textbf{0.719}   &\textbf{0.662}   &\textbf{0.500}    &\textbf{0.526} \\ \hline\hline
\toprule%
& \multicolumn{7}{@{}c@{}}{F-Score $\uparrow$}  \\\cmidrule{2-8}%
                    Method                                                   &Deer   &DNA    &Box     &Eevee    &Polarbear    &Robot   &Skull \\ \hline 
\midrule
				    $\texttt{Time-max}$                                      &0.366  &0.424  &0.364   &0.300    &0.208        &0.298   &0.303 \\ \hline
					$\texttt{DnCNN-S}$ \cite{zhang2017beyond}                &0.379  &0.367	 &0.353	  &0.409	&0.321	      &0.381   &0.336  \\ \hline
					$\texttt{RED}$ \cite{mao2016image}                       &0.368	 &0.606	 &0.541	  &0.338	&0.343	      &0.357   &0.347  \\ \hline
					$\texttt{NBNet}$ \cite{cheng2021nbnet}                   &0.476	 &0.278	 &0.506	  &0.346	&0.268	      &0.314   &0.381     \\ \hline
					$\texttt{U-Net}$ \cite{ronneberger2015u}                 &0.403	 &0.471	 &0.243	  &0.378	&0.282	      &0.292   &0.306  \\ \hline
				% 	$\texttt{U-Net}_\texttt{MS}$                              &22.46	 &25.05  &20.81	  &20.34    &19.86        &20.64   &19.43  \\ \hline 
					\texttt{SARNet} (Ours)         &0.593	&0.704	&0.502	&0.330	&0.370	&0.363	&0.506 \\
					\hline
					$\texttt{SARNet}_{\texttt{MV}}$ (Ours)  & \textbf{0.605}	&\textbf{0.715}	&\textbf{0.574}	&\textbf{0.410}	&\textbf{0.391}	&\textbf{0.400}	&\textbf{0.613} \\ \hline
\toprule%
& \multicolumn{7}{@{}c@{}}{Chamfer distance $\downarrow$}  \\\cmidrule{2-8}%
                    Method                                                   &Deer   &DNA    &Box     &Eevee    &Polarbear    &Robot   &Skull \\ \hline 
\midrule
				    $\texttt{Time-max}$                                      &0.016	&0.012	&0.019	&0.022	&0.091	&0.023	&0.022 \\ \hline
					$\texttt{DnCNN-S}$ \cite{zhang2017beyond}                &0.020	&0.014	&0.027	&0.018	&0.022	&0.018	&0.022  \\ \hline
					$\texttt{RED}$ \cite{mao2016image}                       &0.018	&0.009	&0.013	&0.018	&0.021	&0.018	&0.021  \\ \hline
					$\texttt{NBNet}$ \cite{cheng2021nbnet}                   &0.016	&0.020	&0.013	&0.020	&0.025	&0.021	&0.021     \\ \hline
					$\texttt{U-Net}$ \cite{ronneberger2015u}                 &0.018	&0.012	&0.031	&0.018	&0.022	&0.024	&0.020  \\ \hline
				% 	$\texttt{U-Net}_\texttt{MS}$                              &22.46	 &25.05  &20.81	  &20.34    &19.86        &20.64   &19.43  \\ \hline 
					\texttt{SARNet} (Ours)         &0.011	&0.008	&0.021	&0.024	&0.018	&0.018	&0.012 \\
					\hline
					$\texttt{SARNet}_{\texttt{MV}}$  (Ours) &\textbf{0.011}	&\textbf{0.008}	&\textbf{0.011}	&\textbf{0.016}	&\textbf{0.016}	&\textbf{0.018}	&\textbf{0.010} \\ \hline
\botrule
\end{tabular*}
\end{minipage}
\end{center}
%\label{tab:3d}
\end{table}

% 	\begin{table}
% 		\caption{Quantitative comparison (PSNR, SSIM, and LPIPS) of THz image restoration performances for \textbf{Deer} and \textbf{DNA} with different settings. $\uparrow$: higher is better}
% 		\begin{center}
% 		    \vspace{-0.20in}
% 			\begin{scriptsize}
% 			\scriptsize
% 				\begin{tabular}{|c|c|c||c|c||c|c|} \hline
% 				    Method & \multicolumn{2}{c||}{PSNR$\uparrow$} & \multicolumn{2}{c||}{SSIM$\uparrow$} & \multicolumn{2}{c|}{LPIPS$\downarrow$}\\
% 				    \cline{2-3} \cline{4-5} \cline{6-7}
% 					                                   &Deer     &DNA    &Deer   &DNA      &Deer   &DNA \\ \hline\hline
% 					base-Unet \cite{ronneberger2015u}  &19.84	 &25.63  &0.55   &0.78     &0.77   &0.11\\ \hline
% 					Amp-Unet w/o attention             &22.05	 &25.84  &0.80   &0.83     &0.14   &0.08\\ \hline
%                     Phase-Unet w/o attention           &21.14	 &24.98  &0.82   &0.72     &0.14   &0.09\\ \hline
%                     Mix-Unet w/o attention             &21.44    &25.78  &0.81   &0.81     &0.14   &0.08\\ \hline
%                     Amp-Unet w/ attention              &20.97    &26.32  &0.84   &0.90     &0.14   &0.10  \\ \hline
%                     Phase-Unet w/ attention            &22.66    & 25.52 &0.83   &0.86     &0.14   &0.09  \\ \hline
% 					S3AF-Unet (Ours)                   &\textbf{22.84}	 &\textbf{26.84}  &\textbf{0.84}  &\textbf{0.91}   &\textbf{0.14}  &\textbf{0.07}\\ \hline
% 				\end{tabular}
% 			\end{scriptsize}
% 		\end{center}
% 		\label{tab:t2}
%  		\vspace{-0.20in}
% 	\end{table}

\subsection{Ablation Studies}
% We first show that multi-dominant-spectral information could significantly improve restoration performance. 
To verify the  effectiveness of multi-spectral feature fusion, we evaluate the restoration performances with our $\texttt{SARNet}_{\texttt{MV}}$ under different settings  in Table\,\ref{table:ablation_study}. The compared methods include (1) $\texttt{U-Net}$ using a single channel of data (\texttt{Time-max}) without using features of multi-spectral bands; (2) \texttt{U-Net+Amp w/o SAFM} employing multi-band amplitude feature (without the SAFM mechanism) in each of the four spatial-scale branches, except for the finest scale  (that accepts the \texttt{Time-max} image as the input), where  12 spectral bands of amplitude (3 bands/scale) are fed into the four spatial-scale branches with the assignment of the highest-frequency band to the coarsest scale, and vice versa; (3) \texttt{U-Net+Phase w/o SAFM} employing multi-spectral phase features with the same spectral arrangements as (2), and without the SAFM mechanism; (4) \texttt{U-Net+Amp with SAFM} utilizing subspace-attention-guided  multi-spectral amplitude features with  the same spectral arrangements as specified in (2); (5) \texttt{U-Net+Phase with SAFM} utilizing subspace-attention-guided multi-spectral phase features with the same spectral arrangements as in (2); (6) \texttt{SARNet w/o SAFM} concatenating multi-spectral amplitude and phase features (without SAFM) in each of the four spatial-scale branches, except for the finest scale (that accepts the Time-max image as the input), where  totally 24 additional spectral bands of amplitude and phase (3 amplitude plus 3 phase bands for each scale) are fed into the four branches;(7)  $\texttt{SARNet}$ w/o $\texttt{Proj}$ using SAFM to fuse intra-view multi-spectral amplitude and phase features without the aid of subspace projection; (8)  $\texttt{SARNet}_{\texttt{MV}}$ w/o $\texttt{SAFM}$ employing multi-view fusion and multi-spectral amplitude and phase features with the same spectral arrangements as (6) but without subspace-attention guidance; (9) \texttt{SARNet} utilizing intra-view multi-spectral amplitude and phase features with subspace-attention guidance, but without utilizing the inter-view redundancies; (10) $\texttt{SARNet}_{\texttt{MV}}$ utilizing full set of intra-view and inter-view features.

The results clearly demonstrate that the proposed SAFM can well fuse  the spectral features of both amplitude and phase with different characteristics for THz image restoration. 
Specifically, employing additional multi-spectral features of either amplitude or phase as the input of the multi-scale branches in the network (\ie \texttt{U-Net+Amp w/o SAFM} or \texttt{U-Net+Phase w/o SAFM}) can achieve performance improvement over $\texttt{U-Net}$.  Combining both the amplitude and phase features without the proposed subspace-attention-guided fusion (\ie \texttt{SARNet w/o SAFM}) does not outperform \texttt{U-Net+Amp w/o SAFM} and usually leads to worse performances. The main reason is that the characteristics of the amplitude and phase features are too different to be fused to extract useful features with direct fusion methods. This motivates our subspace-attention-guided fusion scheme, which learns to effectively identify and fuse important and complementary features on common ground. The individual impacts of the subspace projection-guided fusion  and the attention-guided fusion can be assessed by checking the performance differences among \texttt{SARNet}, $\texttt{SARNet}$ w/o $\texttt{Proj}$, and $\texttt{SARNet}$ w/o $\texttt{SAFM}$. Furthermore, the multi-view based  $\texttt{SARNet}_{\texttt{MV}}$ can further improve performance by utilizing additional inter-view redundancies, especially on objects with more details such as \textbf{Deer}, \textbf{Box}, and \textbf{Skull}. These results show that the proposed modules all stably achieve performance gains individually and collectively.

\begin{table}
\begin{center}
\begin{minipage}{\textwidth}
\caption{Ablation study in terms of PSNR of THz image restoration performances on \textbf{Deer}, \textbf{DNA}, \textbf{Box},  \textbf{Eevee}, \textbf{Polarbear}, \textbf{Robot}, and \textbf{Skull} with the different variants based on  different settings. ($\uparrow$: higher is better)}
\label{table:ablation_study}
\begin{tabular*}{\textwidth}{@{\extracolsep{\fill}}lccccccc@{\extracolsep{\fill}}}
\toprule%
& \multicolumn{7}{@{}c@{}}{PSNR $\uparrow$}  \\\cmidrule{2-8}%
Method   &Deer   &DNA    &Box     &Eevee    &Polarbear    &Robot   &Skull \\ \hline 
\midrule
					$\texttt{U-Net}$ \cite{ronneberger2015u}  &19.84	 &24.15  &19.77  &19.95  &19.09        &18.80   &17.49    \\ \hline
					$\texttt{U-Net+Amp}$ w/o $\texttt{SAFM}$           &22.05	 &25.84  &20.32  &20.21  &20.48        &20.63   &20.70    \\ \hline
                    $\texttt{U-Net+Phase}$ w/o $\texttt{SAFM}$         &21.14	 &24.98  &20.42  &20.26  &20.15        &20.58   &21.36    \\ \hline

                    $\texttt{U-Net+Amp}$ w/ $\texttt{SAFM}$            &20.97  &26.00  &21.83  &20.22  &20.30        &21.11   &20.18    \\ \hline
                    $\texttt{U-Net+Phase}$ w/ $\texttt{SAFM}$          &22.66  &25.52  &21.65  &20.63  &20.18        &21.50   &21.42    \\ \hline
                    $\texttt{SARNet}$ w/o $\texttt{SAFM}$           &21.44  &25.78  &20.00  &20.32  &20.44        &21.12   &21.18    \\ \hline
                    $\texttt{SARNet}$ w/o $\texttt{Proj}$             &22.40  &25.86      &21.43  &20.46  &20.88        &22.34      &21.87     \\ \hline
                    $\texttt{SARNet}_{\texttt{MV}}$ w/o $\texttt{SAFM}$      &22.49  &25.78     &22.10  &19.91  &20.96        &21.75   &22.47    \\ \hline
					$\texttt{SARNet}$ (Ours)         &\underline{22.98}	 &\underline{26.05}  &\underline{22.67}  &\underline{20.87}   &\underline{21.42}   &\underline{22.66}    &\underline{22.48} 
					\\ \hline
					$\texttt{SARNet}_{\texttt{MV}}$ (Ours)  &\textbf{23.17}	 &\textbf{26.19}  &\textbf{23.23}  &\textbf{20.97}   &\textbf{21.55}   &\textbf{22.68}    &\textbf{23.05} \\ \hline
\botrule

\end{tabular*}
\end{minipage}
\end{center}
\end{table}

		\begin{table}[!]
		\caption{Comparison of the model complexity (the numbers of Parameters, GFLOPs, and run-time) with different methods. Run-time are measured with the Nvidia Titan 2080 Ti. }
		\begin{center}
		  %  \vspace{-0.20in}
			\begin{scriptsize}
			\scriptsize
			\scalebox{1.0}{
				\begin{tabular}{|c|c|c|c|} \hline

					                       Method                             &Params (M)  & GFLOPs   & Run-time (ms)    \\ \hline\hline
					$\texttt{DnCNN-S}$ \cite{zhang2017beyond}                 &0.55        &4.55        &6  \\ \hline
					$\texttt{RED}$ \cite{mao2016image}                        &0.66	       &1.36        &4  \\ \hline
					$\texttt{NBNet}$ \cite{cheng2021nbnet}                    &13.31       &22.20       &25     \\ \hline
					$\texttt{U-Net}$ \cite{ronneberger2015u}    &9.5	       &3.88        &11  \\ \hline
				 	$\texttt{SARNet}$ (Ours)                                  &3.0	       &1.91        &19  \\ \hline 
					$\texttt{SARNet}_{\texttt{MV}}$ (Ours)                    &3.6	       &4.47        &53   \\ \hline
				\end{tabular}}
			\end{scriptsize}
		\end{center}
		\label{table:complexity}
%  		\vspace{-0.4in}
	\end{table}

% 		\begin{table}
% 		\caption{Comparison of the model complexity (the numbers of Parameters and GFLOPs) with different methods. Run-time are measured with the Nvidia Titan 2080 Ti. }
% 		\begin{center}
% 		  %  \vspace{-0.20in}
% 			\begin{scriptsize}
% 			\scriptsize
% 			\scalebox{1.0}{
% 				\begin{tabular}{|c|c|c|} \hline

% 					                       Method                             &Params (M)  & GFLOPs       \\ \hline\hline
% 					$\texttt{DnCNN-S}$ \cite{zhang2017beyond}                 &0.55        &4.55         \\ \hline
% 					$\texttt{RED}$ \cite{mao2016image}                        &0.66	       &1.36          \\ \hline
% 					$\texttt{NBNet}$ \cite{cheng2021nbnet}                    &13.31       &22.20           \\ \hline
% 					$\texttt{U-Net}_\texttt{base}$ \cite{ronneberger2015u}    &9.5	       &3.88          \\ \hline
% 				 	$\texttt{SARNet}$ (Ours)                                  &3.0	       &1.91     \\ \hline 
% 					$\texttt{SARNet}_{\texttt{MV}}$ (Ours)                    &3.6	       &4.47       \\ \hline
% 				\end{tabular}}
% 			\end{scriptsize}
% 		\end{center}
% 		\label{tab:comp}
% %  		\vspace{-0.4in}
% 	\end{table}

% \vspace{-0.1in}

\subsection{Model Complexity}
Table\,\ref{table:complexity} compares the model complexities of the six methods. When compared to the state-of-the-art method $\texttt{NBNet}$~\cite{cheng2021nbnet}, and \texttt{U-Net}, our  $\texttt{SARNet}$ requires  a much fewer number of parameters and GFLOPs. The run-time with \texttt{SARNet} is also less than $\texttt{NBNet}$, but more than $\texttt{U-Net}$ though. In contrast to $\texttt{SARNet}$, $\texttt{SARNet}_{\texttt{MV}}$ achieves the best visual performance while introducing additional computation and storage costs since it involves an additional stage of $\texttt{SARNet}$ restoration, thereby doubling the computation.  All the above comparisons demonstrate that both $\texttt{SARNet}$ and  $\texttt{SARNet}_{\texttt{MV}}$ are promising solutions, considering their much better THz image restoration performances and reasonable computation and memory costs.

\subsection{Limitations}
\texttt{SARNET} uses multi-spectral amplitude/phase data to retrieve geometric information. Depending on the selected THz frequency bands and their SNR, the diffraction-limited system resolution can theoretically push down to ~0.1mm. As water/metal are highly absorptive/reflective materials to the THz wave, our system is not applicable to the aqueous objects or objects hidden inside metallic packages.

Besides, limited by using a single THz source-detector pair, our THz-TDS system operates by a raster scanning approach. Although such a scanning approach makes it still far from real-time applications and is limited to static scenes, there are variants of THz-TDS systems that feature much shorter imaging time. For example, in \cite{li202063}, an $N$-pixel ($N = 63$) THz detector array is developed to offer $N$ times faster image acquisition speed by spreading the THZ light to the detector array.

\section{Conclusions and Future Work}
\label{sec:conclusion}

Aiming at making the invisible visible, we proposed a 3-D THz tomographic imaging system based on multi-view multi-scale spatio-spectral   feature fusion. Considering the physical characteristics of THz waves passing through different materials, our THz imaging methods learn to extracting most predominant spectral  features in different spatial scales for restoring corrupted THz images. The extracted multi-spectral features are then fused on a common ground by the proposed subspace-attention guided fusion and then used to restore THz images in a fine-to-coarse manner.  As a result, the 3D tomography of an object can be reconstructed from the restored 2D THz images  by inverse Radon transform for object  inspection and exploration. Besides intra-view fusion, we have also proposed an inter-view fusion approach to further improve the restoration/reconstruction performance. Our experimental results have confirmed a performance leap from the relevant state-of-the-art techniques in the area.  We believe our findings in this work will shed on light on physics-guided THz computational imaging with advanced computer vision techniques.

As the THz computational imaging research in the computer vision community is still in its early stage, there are several possible directions worth further exploration. From the THz imaging quality point of view, an end-to-end learning framework for direct reconstruction of 3D geometry can avoid the artifacts caused by the typical tomographic reconstruction by the filtered backprojection of 2D projection views, thereby enhancing 3D reconstruction quality further. To this end, it would require to explore a newly designed learning framework involving network models, loss functions, and datasets. Moreover, incorporating the THz beam propagation 3D profile with the deconvolution techniques can further improve THz imaging quality. 
To extend the applications of THz imaging, by leveraging the prior knowledge of light-matter interaction in the THz range, the extension of THz computational imaging to functional imaging of multi-material objects can also be explored.   Last but not least, integrating a massive THz detector array with the THz-TDS system would pave the way to achieve real-time THz tomographic imaging.

\bibliography{terahertz}

%% BioMed_Central_Bib_Style_v1.01

\begin{thebibliography}{68}
% BibTex style file: bmc-mathphys.bst (version 2.1), 2014-07-24
\ifx \bisbn   \undefined \def \bisbn  #1{ISBN #1}\fi
\ifx \binits  \undefined \def \binits#1{#1}\fi
\ifx \bauthor  \undefined \def \bauthor#1{#1}\fi
\ifx \batitle  \undefined \def \batitle#1{#1}\fi
\ifx \bjtitle  \undefined \def \bjtitle#1{#1}\fi
\ifx \bvolume  \undefined \def \bvolume#1{\textbf{#1}}\fi
\ifx \byear  \undefined \def \byear#1{#1}\fi
\ifx \bissue  \undefined \def \bissue#1{#1}\fi
\ifx \bfpage  \undefined \def \bfpage#1{#1}\fi
\ifx \blpage  \undefined \def \blpage #1{#1}\fi
\ifx \burl  \undefined \def \burl#1{\textsf{#1}}\fi
\ifx \doiurl  \undefined \def \doiurl#1{\url{https://doi.org/#1}}\fi
\ifx \betal  \undefined \def \betal{\textit{et al.}}\fi
\ifx \binstitute  \undefined \def \binstitute#1{#1}\fi
\ifx \binstitutionaled  \undefined \def \binstitutionaled#1{#1}\fi
\ifx \bctitle  \undefined \def \bctitle#1{#1}\fi
\ifx \beditor  \undefined \def \beditor#1{#1}\fi
\ifx \bpublisher  \undefined \def \bpublisher#1{#1}\fi
\ifx \bbtitle  \undefined \def \bbtitle#1{#1}\fi
\ifx \bedition  \undefined \def \bedition#1{#1}\fi
\ifx \bseriesno  \undefined \def \bseriesno#1{#1}\fi
\ifx \blocation  \undefined \def \blocation#1{#1}\fi
\ifx \bsertitle  \undefined \def \bsertitle#1{#1}\fi
\ifx \bsnm \undefined \def \bsnm#1{#1}\fi
\ifx \bsuffix \undefined \def \bsuffix#1{#1}\fi
\ifx \bparticle \undefined \def \bparticle#1{#1}\fi
\ifx \barticle \undefined \def \barticle#1{#1}\fi
\bibcommenthead
\ifx \bconfdate \undefined \def \bconfdate #1{#1}\fi
\ifx \botherref \undefined \def \botherref #1{#1}\fi
\ifx \url \undefined \def \url#1{\textsf{#1}}\fi
\ifx \bchapter \undefined \def \bchapter#1{#1}\fi
\ifx \bbook \undefined \def \bbook#1{#1}\fi
\ifx \bcomment \undefined \def \bcomment#1{#1}\fi
\ifx \oauthor \undefined \def \oauthor#1{#1}\fi
\ifx \citeauthoryear \undefined \def \citeauthoryear#1{#1}\fi
\ifx \endbibitem  \undefined \def \endbibitem {}\fi
\ifx \bconflocation  \undefined \def \bconflocation#1{#1}\fi
\ifx \arxivurl  \undefined \def \arxivurl#1{\textsf{#1}}\fi
\csname PreBibitemsHook\endcsname

%%% 1
\bibitem{kamruzzaman2011application}
\begin{barticle}
\bauthor{\bsnm{Kamruzzaman}, \binits{M.}},
\bauthor{\bsnm{ElMasry}, \binits{G.}},
\bauthor{\bsnm{Sun}, \binits{D.-W.}},
\bauthor{\bsnm{Allen}, \binits{P.}}:
\batitle{Application of {NIR} hyperspectral imaging for discrimination of lamb
  muscles}.
\bjtitle{J. Food Engineer.}
\bvolume{104}(\bissue{3}),
\bfpage{332}--\blpage{340}
(\byear{2011})
\end{barticle}
\endbibitem

%%% 2
\bibitem{rotermund1991methods}
\begin{barticle}
\bauthor{\bsnm{Rotermund}, \binits{H.H.}},
\bauthor{\bsnm{Engel}, \binits{W.}},
\bauthor{\bsnm{Jakubith}, \binits{S.}},
\bauthor{\bsnm{Von~Oertzen}, \binits{A.}},
\bauthor{\bsnm{Ertl}, \binits{G.}}:
\batitle{Methods and application of {UV} photoelectron microscopy in
  heterogenous catalysis}.
\bjtitle{Ultramicroscopy}
\bvolume{36}(\bissue{1-3}),
\bfpage{164}--\blpage{172}
(\byear{1991})
\end{barticle}
\endbibitem

%%% 3
\bibitem{yujiri2003passive}
\begin{barticle}
\bauthor{\bsnm{Yujiri}, \binits{L.}},
\bauthor{\bsnm{Shoucri}, \binits{M.}},
\bauthor{\bsnm{Moffa}, \binits{P.}}:
\batitle{Passive millimeter wave imaging}.
\bjtitle{IEEE Microwave Mag.}
\bvolume{4}(\bissue{3}),
\bfpage{39}--\blpage{50}
(\byear{2003})
\end{barticle}
\endbibitem

%%% 4
\bibitem{abbas2021classification}
\begin{barticle}
\bauthor{\bsnm{Abbas}, \binits{A.}},
\bauthor{\bsnm{Abdelsamea}, \binits{M.}},
\bauthor{\bsnm{Gaber}, \binits{M.M.}}:
\batitle{Classification of {COVID-19} in chest {X-ray} images using {DeTraC}
  deep convolutional neural network}.
\bjtitle{Appl. Intell.}
\bvolume{51}(\bissue{2}),
\bfpage{854}--\blpage{864}
(\byear{2021})
\end{barticle}
\endbibitem

%%% 5
\bibitem{round2005preliminary}
\begin{barticle}
\bauthor{\bsnm{Round}, \binits{A.R.}},
\bauthor{\bsnm{Wilkinson}, \binits{S.J.}},
\bauthor{\bsnm{Hall}, \binits{C.J.}},
\bauthor{\bsnm{Rogers}, \binits{K.D.}},
\bauthor{\bsnm{Glatter}, \binits{O.}},
\bauthor{\bsnm{Wess}, \binits{T.}},
\bauthor{\bsnm{Ellis}, \binits{I.O.}}:
\batitle{A preliminary study of breast cancer diagnosis using laboratory based
  small angle x-ray scattering}.
\bjtitle{Physics in Medicine \& Biology}
\bvolume{50}(\bissue{17}),
\bfpage{4159}
(\byear{2005})
\end{barticle}
\endbibitem

%%% 6
\bibitem{tuan2018dental}
\begin{barticle}
\bauthor{\bsnm{Tuan}, \binits{T.M.}},
\bauthor{\bsnm{Fujita}, \binits{H.}},
\bauthor{\bsnm{Dey}, \binits{N.}},
\bauthor{\bsnm{Ashour}, \binits{A.S.}},
\bauthor{\bsnm{Ngoc}, \binits{T.N.}},
\bauthor{\bsnm{Chu}, \binits{D.-T.}}:
\batitle{Dental diagnosis from {X-ray} images: {A}n expert system based on
  fuzzy computing}.
\bjtitle{Biomed. Signal Process. Control}
\bvolume{39},
\bfpage{64}--\blpage{73}
(\byear{2018})
\end{barticle}
\endbibitem

%%% 7
\bibitem{xie2008review}
\begin{botherref}
\oauthor{\bsnm{Xie}, \binits{X.}}:
A review of recent advances in surface defect detection using texture analysis
  techniques.
ELCVIA: Electron. Lett. Comput. Vis. Image Ana.,
1--22
(2008)
\end{botherref}
\endbibitem

%%% 8
\bibitem{de2004risk}
\begin{barticle}
\bauthor{\bparticle{de} \bsnm{Gonzalez}, \binits{A.B.}},
\bauthor{\bsnm{Darby}, \binits{S.}}:
\batitle{Risk of cancer from diagnostic {X-rays}: {E}stimates for the uk and 14
  other countries}.
\bjtitle{The Lancet}
\bvolume{363}(\bissue{9406}),
\bfpage{345}--\blpage{351}
(\byear{2004})
\end{barticle}
\endbibitem

%%% 9
\bibitem{abraham2010non}
\begin{barticle}
\bauthor{\bsnm{braham}, \binits{E.}},
\bauthor{\bsnm{Younus}, \binits{A.}},
\bauthor{\bsnm{Delagnes}, \binits{T.-C.}},
\bauthor{\bsnm{Mounaix}, \binits{P.}}:
\batitle{Non-invasive investigation of art paintings by terahertz imaging}.
\bjtitle{Appl. Physics A}
\bvolume{100}(\bissue{3}),
\bfpage{585}--\blpage{590}
(\byear{2010})
\end{barticle}
\endbibitem

%%% 10
\bibitem{yu2012potential}
\begin{barticle}
\bauthor{\bsnm{Calvin}, \binits{Y.}},
\bauthor{\bsnm{Shuting}, \binits{F.}},
\bauthor{\bsnm{Yiwen}, \binits{S.}},
\bauthor{\bsnm{Emma}, \binits{P.-M.}}:
\batitle{The potential of terahertz imaging for cancer diagnosis: {A} review of
  investigations to date}.
\bjtitle{Quantitative imaging in medicine and surgery}
\bvolume{2}(\bissue{1}),
\bfpage{33}
(\byear{2012})
\end{barticle}
\endbibitem

%%% 11
\bibitem{kim2018rgbd}
\begin{bchapter}
\bauthor{\bsnm{Kim}, \binits{J.}},
\bauthor{\bsnm{Lim}, \binits{H.}},
\bauthor{\bsnm{Ahn}, \binits{S.C.}},
\bauthor{\bsnm{Lee}, \binits{S.}}:
\bctitle{{RGBD} camera based material recognition via surface roughness
  estimation}.
In: \bbtitle{Proc. IEEE Winter Conf. Appl. Comput. Vis.},
pp. \bfpage{1963}--\blpage{1971}
(\byear{2018})
\end{bchapter}
\endbibitem

%%% 12
\bibitem{nunes2020polarization}
\begin{barticle}
\bauthor{\bsnm{Nunes-Pereira}, \binits{E.}},
\bauthor{\bsnm{Peixoto}, \binits{H.}},
\bauthor{\bsnm{Teixeira}, \binits{J.}},
\bauthor{\bsnm{Santos}, \binits{J.}}:
\batitle{Polarization-coded material classification in automotive {LIDAR}
  aiming at safer autonomous driving implementations}.
\bjtitle{Applied Optics}
\bvolume{59}(\bissue{8}),
\bfpage{2530}--\blpage{2540}
(\byear{2020})
\end{barticle}
\endbibitem

%%% 13
\bibitem{clarke1995mri}
\begin{barticle}
\bauthor{\bsnm{Clarke}, \binits{L.}},
\bauthor{\bsnm{Velthuizen}, \binits{R.}},
\bauthor{\bsnm{Camacho}, \binits{M.}},
\bauthor{\bsnm{Heine}, \binits{J.}},
\bauthor{\bsnm{Vaidyanathan}, \binits{M.}},
\bauthor{\bsnm{Hall}, \binits{L.}},
\bauthor{\bsnm{Thatcher}, \binits{R.}},
\bauthor{\bsnm{Silbiger}, \binits{M.}}:
\batitle{{MRI} segmentation: methods and applications}.
\bjtitle{Magnetic Resonance Imag.}
\bvolume{13}(\bissue{3}),
\bfpage{343}--\blpage{368}
(\byear{1995})
\end{barticle}
\endbibitem

%%% 14
\bibitem{chapman1997diffraction}
\begin{barticle}
\bauthor{\bsnm{Chapman}, \binits{D.}},
\bauthor{\bsnm{Homlinson}, \binits{W.}},
\bauthor{\bsnm{Johnston}, \binits{R.}},
\bauthor{\bsnm{Washburn}, \binits{D.}},
\bauthor{\bsnm{Pisano}, \binits{E.}},
\bauthor{\bsnm{Gm{\"u}r}, \binits{N.}},
\bauthor{\bsnm{Zhong}, \binits{Z.}},
\bauthor{\bsnm{Menk}, \binits{R.}},
\bauthor{\bsnm{Arfelli}, \binits{F.}},
\bauthor{\bsnm{Sayers}, \binits{D.}}:
\batitle{Diffraction enhanced {X-ray} imaging}.
\bjtitle{J. Physics in Med. \& Biol.}
\bvolume{42}(\bissue{11}),
\bfpage{2015}
(\byear{1997})
\end{barticle}
\endbibitem

%%% 15
\bibitem{fitzgerald2000phase}
\begin{barticle}
\bauthor{\bsnm{Fitzgerald}, \binits{R.}}:
\batitle{Phase-sensitive {X-ray} imaging}.
\bjtitle{Physics Today}
\bvolume{53}(\bissue{7}),
\bfpage{23}--\blpage{26}
(\byear{2000})
\end{barticle}
\endbibitem

%%% 16
\bibitem{sakdinawat2010nanoscale}
\begin{barticle}
\bauthor{\bsnm{akdinawat}, \binits{A.}},
\bauthor{\bsnm{Attwood}, \binits{D.}}:
\batitle{Nanoscale {X-ray} imaging}.
\bjtitle{Nature Photon.}
\bvolume{4}(\bissue{12}),
\bfpage{840}
(\byear{2010})
\end{barticle}
\endbibitem

%%% 17
\bibitem{cloetens1996phase}
\begin{barticle}
\bauthor{\bsnm{Cloetens}, \binits{P.}},
\bauthor{\bsnm{arrett}, \binits{R.}},
\bauthor{\bsnm{Baruchel}, \binits{J.}},
\bauthor{\bsnm{Guigay}, \binits{J.-P.}},
\bauthor{\bsnm{Schlenker}, \binits{M.}}:
\batitle{Phase objects in synchrotron radiation hard {X}-ray imaging}.
\bjtitle{J. Physics D: Appl. Physics}
\bvolume{29}(\bissue{1}),
\bfpage{133}
(\byear{1996})
\end{barticle}
\endbibitem

%%% 18
\bibitem{peterson2001x}
\begin{barticle}
\bauthor{\bsnm{Peterson}, \binits{J.}},
\bauthor{\bsnm{Paerels}, \binits{F.}},
\bauthor{\bsnm{Kaastra}, \binits{J.}},
\bauthor{\bsnm{Arnaud}, \binits{M.}},
\bauthor{\bsnm{Reiprich}, \binits{T.}},
\bauthor{\bsnm{Fabian}, \binits{A.}},
\bauthor{\bsnm{Mushotzky}, \binits{R.}},
\bauthor{\bsnm{Jernigan}, \binits{J.}},
\bauthor{\bsnm{Sakelliou}, \binits{I.}}:
\batitle{X-ray imaging-spectroscopy of {Abell} 1835}.
\bjtitle{J. Astronomy \& Astrophysics}
\bvolume{365}(\bissue{1}),
\bfpage{104}--\blpage{109}
(\byear{2001})
\end{barticle}
\endbibitem

%%% 19
\bibitem{saeedkia2013handbook}
\begin{botherref}
\oauthor{\bsnm{Saeedkia}, \binits{D.}}:
Handbook of terahertz technology for imaging, sensing and communications.
Woodhead Publishing: Cambridge,
542--578
(2013)
\end{botherref}
\endbibitem

%%% 20
\bibitem{mittleman1999recent}
\begin{barticle}
\bauthor{\bsnm{Mittleman}, \binits{D.}},
\bauthor{\bsnm{Gupta}, \binits{M.}},
\bauthor{\bsnm{Neelamani}, \binits{R.}},
\bauthor{\bsnm{Baraniuk}, \binits{R.}},
\bauthor{\bsnm{Rudd}, \binits{J.}},
\bauthor{\bsnm{Koch}, \binits{M.}}:
\batitle{Recent advances in terahertz imaging}.
\bjtitle{Appl. Physics B}
\bvolume{68}(\bissue{6}),
\bfpage{1085}--\blpage{1094}
(\byear{1999})
\end{barticle}
\endbibitem

%%% 21
\bibitem{jansen2010terahertz}
\begin{barticle}
\bauthor{\bsnm{Jansen}, \binits{C.}},
\bauthor{\bsnm{Wietzke}, \binits{S.}},
\bauthor{\bsnm{Peters}, \binits{O.}},
\bauthor{\bsnm{Scheller}, \binits{M.}},
\bauthor{\bsnm{Vieweg}, \binits{N.}},
\bauthor{\bsnm{Salhi}, \binits{M.}},
\bauthor{\bsnm{Krumbholz}, \binits{N.}},
\bauthor{\bsnm{J{\"o}rdens}, \binits{C.}},
\bauthor{\bsnm{Hochrein}, \binits{T.}},
\bauthor{\bsnm{Koch}, \binits{M.}}:
\batitle{Terahertz imaging: applications and perspectives}.
\bjtitle{Appl. Optics}
\bvolume{49}(\bissue{19}),
\bfpage{48}--\blpage{57}
(\byear{2010})
\end{barticle}
\endbibitem

%%% 22
\bibitem{mittleman2018twenty}
\begin{barticle}
\bauthor{\bsnm{Mittleman}, \binits{D.M.}}:
\batitle{Twenty years of terahertz imaging}.
\bjtitle{Optics Express}
\bvolume{26}(\bissue{8}),
\bfpage{9417}--\blpage{9431}
(\byear{2018})
\end{barticle}
\endbibitem

%%% 23
\bibitem{kawase2003non}
\begin{barticle}
\bauthor{\bsnm{Kawase}, \binits{K.}},
\bauthor{\bsnm{Ogawa}, \binits{Y.}},
\bauthor{\bsnm{Watanabe}, \binits{Y.}},
\bauthor{\bsnm{Inoue}, \binits{H.}}:
\batitle{Non-destructive terahertz imaging of illicit drugs using spectral
  fingerprints}.
\bjtitle{Optics Express}
\bvolume{11}(\bissue{20}),
\bfpage{2549}--\blpage{2554}
(\byear{2003})
\end{barticle}
\endbibitem

%%% 24
\bibitem{fukunaga2016thz}
\begin{botherref}
\oauthor{\bsnm{Fukunaga}, \binits{K.}}:
{THz} technology applied to cultural heritage in practice
(2016)
\end{botherref}
\endbibitem

%%% 25
\bibitem{bowman2018pulsed}
\begin{barticle}
\bauthor{\bsnm{Bowman}, \binits{T.}},
\bauthor{\bsnm{Chavez}, \binits{T.}},
\bauthor{\bsnm{Khan}, \binits{K.}},
\bauthor{\bsnm{Wu}, \binits{J.}},
\bauthor{\bsnm{Chakraborty}, \binits{A.}},
\bauthor{\bsnm{Rajaram}, \binits{N.}},
\bauthor{\bsnm{Bailey}, \binits{K.}},
\bauthor{\bsnm{El-Shenawee}, \binits{M.}}:
\batitle{Pulsed terahertz imaging of breast cancer in freshly excised murine
  tumors}.
\bjtitle{J. Biomed. Optics}
\bvolume{23}(\bissue{2}),
\bfpage{026004}
(\byear{2018})
\end{barticle}
\endbibitem

%%% 26
\bibitem{hung2019terahertz}
\begin{bchapter}
\bauthor{\bsnm{Hung}, \binits{Y.-C.}},
\bauthor{\bsnm{Yang}, \binits{S.-H.}}:
\bctitle{Terahertz deep learning computed tomography}.
In: \bbtitle{Proc. Int. Infrad. Milli. THz. Wav.},
pp. \bfpage{1}--\blpage{2}
(\byear{2019}).
\bcomment{IEEE}
\end{bchapter}
\endbibitem

%%% 27
\bibitem{ronneberger2015u}
\begin{bchapter}
\bauthor{\bsnm{Ronneberger}, \binits{O.}},
\bauthor{\bsnm{Fischer}, \binits{P.}},
\bauthor{\bsnm{Brox}, \binits{T.}}:
\bctitle{{U-Net}: {Convolutional} networks for biomedical image segmentation}.
In: \bbtitle{Proc. Int. Conf. Medical Image Comput. Comput.-Assisted
  Intervention},
pp. \bfpage{234}--\blpage{241}
(\byear{2015})
\end{bchapter}
\endbibitem

%%% 28
\bibitem{cheng2021nbnet}
\begin{bchapter}
\bauthor{\bsnm{Cheng}, \binits{S.}},
\bauthor{\bsnm{Wang}, \binits{Y.}},
\bauthor{\bsnm{Huang}, \binits{H.}},
\bauthor{\bsnm{Liu}, \binits{D.}},
\bauthor{\bsnm{Fan}, \binits{H.}},
\bauthor{\bsnm{Liu}, \binits{S.}}:
\bctitle{{NBNet}: {Noise} basis learning for image denoising with subspace
  projection}.
In: \bbtitle{Proc. IEEE/CVF Int. Conf. Comput. Vis. Pattern Recognit.},
pp. \bfpage{4896}--\blpage{4906}
(\byear{2021})
\end{bchapter}
\endbibitem

%%% 29
\bibitem{born2013principles}
\begin{botherref}
\oauthor{\bsnm{Born}, \binits{M.}},
\oauthor{\bsnm{Wolf}, \binits{E.}}:
Principles of optics: {E}lectromagnetic theory of propagation, interference and
  diffraction of light
(2013)
\end{botherref}
\endbibitem

%%% 30
\bibitem{hack2014terahertz}
\begin{barticle}
\bauthor{\bsnm{Hack}, \binits{E.}},
\bauthor{\bsnm{Zolliker}, \binits{P.}}:
\batitle{Terahertz holography for imaging amplitude and phase objects}.
\bjtitle{Optics Express}
\bvolume{22}(\bissue{13}),
\bfpage{16079}--\blpage{16086}
(\byear{2014})
\end{barticle}
\endbibitem

%%% 31
\bibitem{recur2012propagation}
\begin{barticle}
\bauthor{\bsnm{Recur}, \binits{B.}},
\bauthor{\bsnm{Guillet}, \binits{J.-P.}},
\bauthor{\bsnm{Manek-H{\"o}nninger}, \binits{I.}},
\bauthor{\bsnm{Delagnes}, \binits{J.-C.}},
\bauthor{\bsnm{Benharbone}, \binits{W.}},
\bauthor{\bsnm{Desbarats}, \binits{P.}},
\bauthor{\bsnm{Domenger}, \binits{J.-P.}},
\bauthor{\bsnm{Canioni}, \binits{L.}},
\bauthor{\bsnm{Mounaix}, \binits{P.}}:
\batitle{Propagation beam consideration for {3D THz} computed tomography}.
\bjtitle{Optics Express}
\bvolume{20}(\bissue{6}),
\bfpage{5817}--\blpage{5829}
(\byear{2012})
\end{barticle}
\endbibitem

%%% 32
\bibitem{su2021seeing}
\begin{bchapter}
\bauthor{\bsnm{Su}, \binits{W.-T.}},
\bauthor{\bsnm{Hung}, \binits{Y.-C.}},
\bauthor{\bsnm{Yu}, \binits{P.-J.}},
\bauthor{\bsnm{Yang}, \binits{S.-H.}},
\bauthor{\bsnm{Lin}, \binits{C.-W.}}:
\bctitle{Seeing through a black box: {Toward} high-quality terahertz
  tomographic imaging via multi-scale spatio-spectral image fusion}.
In: \bbtitle{Proc. European Conf. Comput. Vis.}
(\byear{2022})
\end{bchapter}
\endbibitem

%%% 33
\bibitem{su2023THzImaging}
\begin{barticle}
\bauthor{\bsnm{Su}, \binits{W.-T.}},
\bauthor{\bsnm{Hung}, \binits{Y.-C.}},
\bauthor{\bsnm{Yu}, \binits{P.-J.}},
\bauthor{\bsnm{Lin}, \binits{C.-W.}},
\bauthor{\bsnm{Yang}, \binits{S.-H.}}:
\batitle{Physics-guided terahertz computational imaging: {A} tutorial on
  sate-of-the-art techniques}.
\bjtitle{IEEE Signal Process. Mag.}
\bvolume{40}(\bissue{2}),
\bfpage{32}--\blpage{45}
(\byear{2023})
\end{barticle}
\endbibitem

%%% 34
\bibitem{zhang2017beyond}
\begin{barticle}
\bauthor{\bsnm{Zhang}, \binits{K.}},
\bauthor{\bparticle{ana} \bsnm{Y.~Chen}, \binits{W.Z.}},
\bauthor{\bsnm{Meng}, \binits{D.}},
\bauthor{\bsnm{Zhang}, \binits{L.}}:
\batitle{Beyond a {Gaussian} denoiser: Residual learning of deep {CNN} for
  image denoising}.
\bjtitle{IEEE Trans. Image Process.}
\bvolume{26}(\bissue{7}),
\bfpage{3142}--\blpage{3155}
(\byear{2017})
\end{barticle}
\endbibitem

%%% 35
\bibitem{mao2016image}
\begin{bchapter}
\bauthor{\bsnm{Mao}, \binits{X.}},
\bauthor{\bsnm{Shen}, \binits{C.}},
\bauthor{\bsnm{Yang}, \binits{Y.-B.}}:
\bctitle{Image restoration using very deep convolutional encoder-decoder
  networks with symmetric skip connections}.
In: \bbtitle{Proc. Adv. Neural Inf. Process. Syst.},
pp. \bfpage{2802}--\blpage{2810}
(\byear{2016})
\end{bchapter}
\endbibitem

%%% 36
\bibitem{zhou2019spatio}
\begin{bchapter}
\bauthor{\bsnm{Zhou}, \binits{S.}},
\bauthor{\bsnm{Zhang}, \binits{J.}},
\bauthor{\bsnm{Pan}, \binits{J.}},
\bauthor{\bsnm{Xie}, \binits{H.}},
\bauthor{\bsnm{Zuo}, \binits{W.}},
\bauthor{\bsnm{Ren}, \binits{J.}}:
\bctitle{Spatio-temporal filter adaptive network for video deblurring}.
In: \bbtitle{Proc. IEEE/CVF Int. Conf. Comput. Vis.},
pp. \bfpage{2482}--\blpage{2491}
(\byear{2019})
\end{bchapter}
\endbibitem

%%% 37
\bibitem{zhang2020residual}
\begin{botherref}
\oauthor{\bsnm{Zhang}, \binits{Y.}},
\oauthor{\bsnm{Tian}, \binits{Y.}},
\oauthor{\bsnm{Kong}, \binits{Y.}},
\oauthor{\bsnm{Zhong}, \binits{B.}},
\oauthor{\bsnm{Fu}, \binits{Y.}}:
Residual dense network for image restoration.
IEEE Trans. Pattern Anal. Mach. Intell.
(2020)
\end{botherref}
\endbibitem

%%% 38
\bibitem{zhang2018ffdnet}
\begin{barticle}
\bauthor{\bsnm{Zhang}, \binits{K.}},
\bauthor{\bsnm{Zuo}, \binits{W.M.}},
\bauthor{\bsnm{Zhang}, \binits{L.}}:
\batitle{{FFDNet}: Toward a fast and flexible solution for cnn-based image
  denoising}.
\bjtitle{IEEE Trans. Image Process.}
\bvolume{27}(\bissue{9}),
\bfpage{4608}--\blpage{4622}
(\byear{2018})
\end{barticle}
\endbibitem

%%% 39
\bibitem{popescu2010point}
\begin{barticle}
\bauthor{\bsnm{Popescu}, \binits{D.C.}},
\bauthor{\bsnm{Ellicar}, \binits{A.D.}}:
\batitle{Point spread function estimation for a terahertz imaging system}.
\bjtitle{EURASIP J. Adv. Signal Process.}
\bvolume{2010}(\bissue{1}),
\bfpage{575817}
(\byear{2010})
\end{barticle}
\endbibitem

%%% 40
\bibitem{popescu2009phantom}
\begin{bchapter}
\bauthor{\bsnm{Popescu}, \binits{D.C.}},
\bauthor{\bsnm{Hellicar}, \binits{A.}},
\bauthor{\bsnm{Li}, \binits{Y.}}:
\bctitle{Phantom-based point spread function estimation for terahertz imaging
  system},
pp. \bfpage{629}--\blpage{639}
(\byear{2009})
\end{bchapter}
\endbibitem

%%% 41
\bibitem{wong2019computational}
\begin{barticle}
\bauthor{\bsnm{Wong}, \binits{T.M.}},
\bauthor{\bsnm{Kahl}, \binits{M.}},
\bauthor{\bsnm{Bol{\'\i}var}, \binits{P.H.}},
\bauthor{\bsnm{Kolb}, \binits{A.}}:
\batitle{Computational image enhancement for frequency modulated continuous
  wave (fmcw) {THz} image}.
\bjtitle{J. Infrar., Millimeter, Terahertz Waves}
\bvolume{40}(\bissue{7}),
\bfpage{775}--\blpage{800}
(\byear{2019})
\end{barticle}
\endbibitem

%%% 42
\bibitem{hung2019kernel}
\begin{bchapter}
\bauthor{\bsnm{Hung}, \binits{Y.-C.}},
\bauthor{\bsnm{Yang}, \binits{S.-H.}}:
\bctitle{Kernel size characterization for deep learning terahertz tomography},
pp. \bfpage{1}--\blpage{2}
(\byear{2019})
\end{bchapter}
\endbibitem

%%% 43
\bibitem{vaswani2017attention}
\begin{botherref}
\oauthor{\bsnm{Vaswani}, \binits{A.}},
\oauthor{\bsnm{Shazeer}, \binits{N.}},
\oauthor{\bsnm{Parmar}, \binits{N.}},
\oauthor{\bsnm{Uszkoreit}, \binits{J.}},
\oauthor{\bsnm{Jones}, \binits{L.}},
\oauthor{\bsnm{Gomez}, \binits{A.N.}},
\oauthor{\bsnm{Kaiser}, \binits{{\L}.}},
\oauthor{\bsnm{Polosukhin}, \binits{I.}}:
Attention is all you need.
Proc. Adv. Neural Inf. Process. Syst.
\textbf{30}
(2017)
\end{botherref}
\endbibitem

%%% 44
\bibitem{dosovitskiy2020image}
\begin{botherref}
\oauthor{\bsnm{Dosovitskiy}, \binits{A.}},
\oauthor{\bsnm{Beyer}, \binits{L.}},
\oauthor{\bsnm{Kolesnikov}, \binits{A.}},
\oauthor{\bsnm{Weissenborn}, \binits{D.}},
\oauthor{\bsnm{Zhai}, \binits{X.}},
\oauthor{\bsnm{Unterthiner}, \binits{T.}},
\oauthor{\bsnm{Dehghani}, \binits{M.}},
\oauthor{\bsnm{Minderer}, \binits{M.}},
\oauthor{\bsnm{Heigold}, \binits{G.}},
\oauthor{\bsnm{Gelly}, \binits{S.}}:
An image is worth 16x16 words: {Transformers} for image recognition at scale.
arXiv preprint arXiv:2010.11929
(2020)
\end{botherref}
\endbibitem

%%% 45
\bibitem{wu2020visual}
\begin{botherref}
\oauthor{\bsnm{Wu}, \binits{B.}},
\oauthor{\bsnm{Xu}, \binits{C.}},
\oauthor{\bsnm{Dai}, \binits{X.}},
\oauthor{\bsnm{Wan}, \binits{A.}},
\oauthor{\bsnm{Zhang}, \binits{P.}},
\oauthor{\bsnm{Yan}, \binits{Z.}},
\oauthor{\bsnm{Tomizuka}, \binits{M.}},
\oauthor{\bsnm{Gonzalez}, \binits{J.}},
\oauthor{\bsnm{Keutzer}, \binits{K.}},
\oauthor{\bsnm{Vajda}, \binits{P.}}:
Visual transformers: {Token}-based image representation and processing for
  computer vision.
arXiv preprint arXiv:2006.03677
(2020)
\end{botherref}
\endbibitem

%%% 46
\bibitem{carion2020end}
\begin{bchapter}
\bauthor{\bsnm{Carion}, \binits{N.}},
\bauthor{\bsnm{Massa}, \binits{F.}},
\bauthor{\bsnm{Synnaeve}, \binits{G.}},
\bauthor{\bsnm{Usunier}, \binits{N.}},
\bauthor{\bsnm{Kirillov}, \binits{A.}},
\bauthor{\bsnm{Zagoruyko}, \binits{S.}}:
\bctitle{End-to-end object detection with transformers}.
In: \bbtitle{Proc. European Conf. Comput. Vis.},
pp. \bfpage{213}--\blpage{229}
(\byear{2020}).
\bcomment{Springer}
\end{bchapter}
\endbibitem

%%% 47
\bibitem{liu2018deep}
\begin{barticle}
\bauthor{\bsnm{Liu}, \binits{F.}},
\bauthor{\bsnm{Jang}, \binits{H.}},
\bauthor{\bsnm{Kijowski}, \binits{R.}},
\bauthor{\bsnm{Bradshaw}, \binits{T.}},
\bauthor{\bsnm{McMillan}, \binits{A.B.}}:
\batitle{Deep learning {MR} imaging-based attenuation correction for {PET/MR}
  imaging}.
\bjtitle{Radiology}
\bvolume{286}(\bissue{2}),
\bfpage{676}--\blpage{684}
(\byear{2018})
\end{barticle}
\endbibitem

%%% 48
\bibitem{chen2021pre}
\begin{bchapter}
\bauthor{\bsnm{Chen}, \binits{H.}},
\bauthor{\bsnm{Wang}, \binits{Y.}},
\bauthor{\bsnm{Guo}, \binits{T.}},
\bauthor{\bsnm{Xu}, \binits{C.}},
\bauthor{\bsnm{Deng}, \binits{Y.}},
\bauthor{\bsnm{Liu}, \binits{Z.}},
\bauthor{\bsnm{Ma}, \binits{S.}},
\bauthor{\bsnm{Xu}, \binits{C.}},
\bauthor{\bsnm{Xu}, \binits{C.}},
\bauthor{\bsnm{Gao}, \binits{W.}}:
\bctitle{Pre-trained image processing transformer}.
In: \bbtitle{Proc. IEEE/CVF Conf. Comput. Vis. Pattern Recognit.},
pp. \bfpage{12299}--\blpage{12310}
(\byear{2021})
\end{bchapter}
\endbibitem

%%% 49
\bibitem{cao2021video}
\begin{botherref}
\oauthor{\bsnm{Cao}, \binits{J.}},
\oauthor{\bsnm{Li}, \binits{Y.}},
\oauthor{\bsnm{Zhang}, \binits{K.}},
\oauthor{\bsnm{Van~Gool}, \binits{L.}}:
Video super-resolution transformer.
arXiv preprint arXiv:2106.06847
(2021)
\end{botherref}
\endbibitem

%%% 50
\bibitem{wang2022uformer}
\begin{bchapter}
\bauthor{\bsnm{Wang}, \binits{Z.}},
\bauthor{\bsnm{Cun}, \binits{X.}},
\bauthor{\bsnm{Bao}, \binits{J.}},
\bauthor{\bsnm{Zhou}, \binits{W.}},
\bauthor{\bsnm{Liu}, \binits{J.}},
\bauthor{\bsnm{Li}, \binits{H.}}:
\bctitle{Uformer: {A} general {U}-shaped transformer for image restoration}.
In: \bbtitle{Proc. IEEE/CVF Conf. Comput. Vis. Pattern Recognit.},
pp. \bfpage{17683}--\blpage{17693}
(\byear{2022})
\end{bchapter}
\endbibitem

%%% 51
\bibitem{kang2017deep}
\begin{barticle}
\bauthor{\bsnm{Kang}, \binits{E.}},
\bauthor{\bsnm{Min}, \binits{J.}},
\bauthor{\bsnm{Ye}, \binits{J.C.}}:
\batitle{A deep convolutional neural network using directional wavelets for
  low-dose {X}-ray {CT} reconstruction}.
\bjtitle{J. Medical physics}
\bvolume{44}(\bissue{10}),
\bfpage{360}--\blpage{375}
(\byear{2017})
\end{barticle}
\endbibitem

%%% 52
\bibitem{jin2017deep}
\begin{barticle}
\bauthor{\bsnm{Jin}, \binits{K.H.}},
\bauthor{\bsnm{McCann}, \binits{M.T.}},
\bauthor{\bsnm{Froustey}, \binits{E.}},
\bauthor{\bsnm{Unser}, \binits{M.}}:
\batitle{Deep convolutional neural network for inverse problems in imaging}.
\bjtitle{IEEE Trans. Image Process.}
\bvolume{26}(\bissue{9}),
\bfpage{4509}--\blpage{4522}
(\byear{2017})
\end{barticle}
\endbibitem

%%% 53
\bibitem{zhu2018image}
\begin{barticle}
\bauthor{\bsnm{Zhu}, \binits{B.}},
\bauthor{\bsnm{Liu}, \binits{J.Z.}},
\bauthor{\bsnm{Cauley}, \binits{S.F.}},
\bauthor{\bsnm{Rosen}, \binits{R.B.}},
\bauthor{\bsnm{S.Rosen}, \binits{M.}}:
\batitle{Image reconstruction by domain-transform manifold learning}.
\bjtitle{Nature}
\bvolume{555}(\bissue{7697}),
\bfpage{487}--\blpage{492}
(\byear{2018})
\end{barticle}
\endbibitem

%%% 54
\bibitem{schultz2001hyperspectral}
\begin{barticle}
\bauthor{\bsnm{Schultz}, \binits{R.}},
\bauthor{\bsnm{Nielsen}, \binits{T.}},
\bauthor{\bsnm{Zavaleta}, \binits{R.J.}},
\bauthor{\bsnm{Wyatt}, \binits{R.}},
\bauthor{\bsnm{Garner}, \binits{H.}}:
\batitle{Hyperspectral imaging: {A} novel approach for microscopic analysis}.
\bjtitle{Cytometry}
\bvolume{43}(\bissue{4}),
\bfpage{239}--\blpage{247}
(\byear{2001})
\end{barticle}
\endbibitem

%%% 55
\bibitem{ozdemir2020deep}
\begin{barticle}
\bauthor{\bsnm{Ozdemir}, \binits{A.}},
\bauthor{\bsnm{Polat}, \binits{K.}}:
\batitle{Deep learning applications for hyperspectral imaging: a systematic
  review}.
\bjtitle{J. Institute Electro. Comput.}
\bvolume{2}(\bissue{1}),
\bfpage{39}--\blpage{56}
(\byear{2020})
\end{barticle}
\endbibitem

%%% 56
\bibitem{geladi2004hyperspectral}
\begin{barticle}
\bauthor{\bsnm{Geladi}, \binits{P.}},
\bauthor{\bsnm{Burger}, \binits{J.}},
\bauthor{\bsnm{Lestander}, \binits{T.}}:
\batitle{Hyperspectral imaging: {calibration} problems and solutions}.
\bjtitle{Chemometrics Intell. Lab. Syst.}
\bvolume{72}(\bissue{2}),
\bfpage{209}--\blpage{217}
(\byear{2004})
\end{barticle}
\endbibitem

%%% 57
\bibitem{dorney2001material}
\begin{barticle}
\bauthor{\bsnm{Dorney}, \binits{T.D.}},
\bauthor{\bsnm{Baraniuk}, \binits{R.G.}},
\bauthor{\bsnm{Mittleman}, \binits{D.M.}}:
\batitle{Material parameter estimation with terahertz time-domain
  spectroscopy}.
\bjtitle{J. Optical Society America A}
\bvolume{18}(\bissue{7}),
\bfpage{1562}--\blpage{1571}
(\byear{2001})
\end{barticle}
\endbibitem

%%% 58
\bibitem{ljubenovic2020cnn}
\begin{bchapter}
\bauthor{\bsnm{Ljubenovic}, \binits{M.}},
\bauthor{\bsnm{Bazrafkan}, \binits{S.}},
\bauthor{\bsnm{Beenhouwer}, \binits{J.D.}},
\bauthor{\bsnm{Sijbers}, \binits{J.}}:
\bctitle{{CNN}-based deblurring of terahertz images.},
pp. \bfpage{323}--\blpage{330}
(\byear{2020})
\end{bchapter}
\endbibitem

%%% 59
\bibitem{wong2019training}
\begin{bchapter}
\bauthor{\bsnm{Wong}, \binits{T.M.}},
\bauthor{\bsnm{Kahl}, \binits{M.}},
\bauthor{\bsnm{Haring-Bol{\'\i}var}, \binits{P.}},
\bauthor{\bsnm{Kolb}, \binits{A.}},
\bauthor{\bsnm{M{\"o}ller}, \binits{M.}}:
\bctitle{Training auto-encoder-based optimizers for terahertz image
  reconstruction},
pp. \bfpage{93}--\blpage{106}
(\byear{2019})
\end{bchapter}
\endbibitem

%%% 60
\bibitem{zhang2019self}
\begin{bchapter}
\bauthor{\bsnm{Zhang}, \binits{H.}},
\bauthor{\bsnm{Goodfellow}, \binits{I.}},
\bauthor{\bsnm{Metaxas}, \binits{D.}},
\bauthor{\bsnm{Odena}, \binits{A.}}:
\bctitle{Self-attention generative adversarial networks}.
In: \bbtitle{Proc. Int. Conf. Mach. Learn.},
pp. \bfpage{7354}--\blpage{7363}
(\byear{2019})
\end{bchapter}
\endbibitem

%%% 61
\bibitem{meyer10matrix}
\begin{bbook}
\bauthor{\bsnm{Meyer}, \binits{C.D.}}:
\bbtitle{Matrix Analysis and Applied Linear Algebra}.
\bpublisher{SIAM},
\blocation{Philadelphia, PA, USA}
(\byear{2000})
\end{bbook}
\endbibitem

%%% 62
\bibitem{qin2020ffa}
\begin{bchapter}
\bauthor{\bsnm{Qin}, \binits{X.}},
\bauthor{\bsnm{Wang}, \binits{X.}},
\bauthor{\bsnm{Bai}, \binits{Y.}},
\bauthor{\bsnm{Xie}, \binits{X.}},
\bauthor{\bsnm{Jia}, \binits{H.}}:
\bctitle{{FFA-Net}: {Feature} fusion attention network for single image
  dehazing}.
In: \bbtitle{Proc. AAAI Conf. on Artif. Intell.},
vol. \bseriesno{34},
pp. \bfpage{11908}--\blpage{11915}
(\byear{2020})
\end{bchapter}
\endbibitem

%%% 63
\bibitem{kak2001algorithms}
\begin{botherref}
\oauthor{\bsnm{Kak}, \binits{A.C.}}:
Algorithms for reconstruction with nondiffracting sources.
Principles of Comput. Tomographic Imag.,
49--112
(2001)
\end{botherref}
\endbibitem

%%% 64
\bibitem{recur2011investigation}
\begin{barticle}
\bauthor{\bsnm{Recur}, \binits{B.}},
\bauthor{\bsnm{Younus}, \binits{A.}},
\bauthor{\bsnm{Salort}, \binits{S.}},
\bauthor{\bsnm{Mounaix}, \binits{P.}},
\bauthor{\bsnm{Chassagne}, \binits{B.}},
\bauthor{\bsnm{Desbarats}, \binits{P.}},
\bauthor{\bsnm{Caumes}, \binits{J.}},
\bauthor{\bsnm{Abraham}, \binits{E.}}:
\batitle{Investigation on reconstruction methods applied to {3D} terahertz
  computed tomography}.
\bjtitle{Optics Express}
\bvolume{19}(\bissue{6}),
\bfpage{5105}--\blpage{5117}
(\byear{2011})
\end{barticle}
\endbibitem

%%% 65
\bibitem{janke2005asynchronous}
\begin{barticle}
\bauthor{\bsnm{Janke}, \binits{C.}},
\bauthor{\bsnm{F{\"o}rst}, \binits{M.}},
\bauthor{\bsnm{Nagel}, \binits{M.}},
\bauthor{\bsnm{Kurz}, \binits{H.}},
\bauthor{\bsnm{Bartels}, \binits{A.}}:
\batitle{Asynchronous optical sampling for high-speed characterization of
  integrated resonant terahertz sensors}.
\bjtitle{Optics Lett.}
\bvolume{30}(\bissue{11}),
\bfpage{1405}--\blpage{1407}
(\byear{2005})
\end{barticle}
\endbibitem

%%% 66
\bibitem{he2015delving}
\begin{bchapter}
\bauthor{\bsnm{He}, \binits{K.}},
\bauthor{\bsnm{Zhang}, \binits{X.}},
\bauthor{\bsnm{Ren}, \binits{S.}},
\bauthor{\bsnm{Sun}, \binits{J.}}:
\bctitle{Delving deep into rectifiers: {Surpassing} human-level performance on
  {ImageNet} classification}.
In: \bbtitle{Proc. IEEE/CVF Int. Conf. Comput. Vis.},
pp. \bfpage{1026}--\blpage{1034}
(\byear{2015})
\end{bchapter}
\endbibitem

%%% 67
\bibitem{xie2020pix2vox}
\begin{barticle}
\bauthor{\bsnm{Xie}, \binits{H.}},
\bauthor{\bsnm{Yao}, \binits{H.}},
\bauthor{},
\bauthor{\bsnm{Zhang}, \binits{S.P.}},
\bauthor{\bsnm{Zhou}, \binits{S.C.}},
\bauthor{\bsnm{Sun}, \binits{W.X.}}:
\batitle{{Pix2Vox++}: {Multi}-scale context-aware {3D} object reconstruction
  from single and multiple images}.
\bjtitle{Int. J. Comput. Vis.}
\bvolume{128}(\bissue{12}),
\bfpage{2919}--\blpage{2935}
(\byear{2020})
\end{barticle}
\endbibitem

%%% 68
\bibitem{li202063}
\begin{bchapter}
\bauthor{\bsnm{Li}, \binits{X.}},
\bauthor{\bsnm{Jarrahi}, \binits{M.}}:
\bctitle{A 63-pixel plasmonic photoconductive terahertz focal-plane array}.
In: \bbtitle{Proc. IEEE/MTT-S Int. Microwave Symp. (IMS)},
pp. \bfpage{91}--\blpage{94}
(\byear{2020})
\end{bchapter}
\endbibitem

\end{thebibliography}
% \bibliography{sn-bibliography}% common bib file
%% if required, the content of .bbl file can be included here once bbl is generated
%%\input sn-article.bbl

%% Default %%
%%\input sn-sample-bib.tex%

\end{document}